\documentclass[12pt]{article}
\usepackage{jheppub}
\usepackage{mathtools}
\usepackage{mathrsfs}
\usepackage{epstopdf}
\usepackage{verbatim}

\def\be{\begin{equation}}
\def\ee{\end{equation}}
\def\bea{\begin{eqnarray}}
\def\eea{\end{eqnarray}}
\def\ba#1\ea{\begin{align}#1\end{align}}
\def\bg#1\eg{\begin{gather}#1\end{gather}}
\def\bm#1\em{\begin{multline}#1\end{multline}}
\def\bmd#1\emd{\begin{multlined}#1\end{multlined}}

\def\b{\beta}

\def\d{\delta}
\def\D{\Delta}
\def\e{\epsilon}

\def\G{\Gamma}

\def\l{\lambda}
\def\L{\Lambda}
\def\m{\mu}
\def\n{\nu}
\def\p{\phi}

\def\r{\rho}
\def\s{\sigma}

\def\la{\label}

\def\re{\ref}
\def\er{\eqref}
\def\se{\section}
\def\sse{\subsection}

\def\fr{\frac}

\def\pa{\partial}
\def\td{\tilde}

\def\ph{\phantom}
\def\eq{\equiv}
\def\wg{\wedge}
\def\nn{\nonumber}
\def\qu{\quad}
\def\qqu{\qquad}
\def\lt{\left}
\def\rt{\right}
\def\({\left(}
\def\){\right)}
\def\[{\left[}
\def\]{\right]}
\def\<{\langle}
\def\>{\rangle}
\def\lra{\leftrightarrow}

\def\cl{{\mathcal L}}
\def\cn{{\mathcal N}}
\def\co{{\mathcal O}}
\def\wb{{\bar w}}
\def\zb{{\bar z}}
\def\tA{{\td A}}
\def\tJ{{\td J}}

\def\CFT{{\rm CFT}}

\newcommand{\T}[3]{{#1^{#2}_{\ph{#2}#3}}}

\def\tq{{\tilde q}}

\def\vy{{\vec y}}
\def\vx{{\vec x}}

\begin{document}

\subheader{SU-ITP-14/23, MIT-CTP/4598}
\title{Explicitly Broken Supersymmetry with\\Exactly Massless Moduli}
\author[a]{Xi Dong,}
\author[a,b]{Daniel Z. Freedman}
\author[a]{and Yue Zhao}
\affiliation[a]{Stanford Institute for Theoretical Physics, Department of Physics, Stanford University, Stanford, CA 94305, U.S.A.}
\affiliation[b]{Center for Theoretical Physics and Department of Mathematics, Massachusetts Institute of Technology, Cambridge, MA 02139, U.S.A.}
\emailAdd{xidong@stanford.edu, dzf@math.mit.edu, zhaoyue@stanford.edu}


\abstract{The AdS/CFT correspondence is applied to an analogue of the little hierarchy problem in three-dimensional supersymmetric theories.  The bulk is governed by a supergravity theory
in which a U(1) $\times$ U(1) R-symmetry is gauged by Chern-Simons fields.  The bulk theory is deformed by a boundary term quadratic in the gauge fields.
  It breaks SUSY completely and sources an exactly marginal operator in the dual CFT.  SUSY breaking is communicated by gauge interactions to  bulk scalar fields and their spinor superpartners. The bulk-to-boundary propagator of the Chern-Simons fields is a total derivative with respect to the bulk coordinates. Integration by parts and the Ward identity permit evaluation of SUSY breaking effects to all orders in the strength of the deformation.  The R-charges of scalars and spinors differ so large SUSY breaking mass shifts are generated.  Masses of R-neutral particles such as scalar moduli are not shifted to any order in the deformation strength, despite the fact that they may couple to R-charged fields running in loops.
 We also obtain a universal deformation formula for correlation functions under an exactly marginal deformation by a product of holomorphic and anti-holomorphic U(1) currents.}

\maketitle

\se{Introduction}

In a quantum field theory, scalar fields
typically have unprotected masses and are naturally heavy due to
quantum corrections.
 Current LHC results  pose
challenges to supersymmetry (SUSY) as the solution of this hierarchy
problem. Although model building or hidden experimental
signatures\footnote{For more details, please see
\cite{Craig:2013cxa,Dimopoulos:2014aua} and the references therein.}
may rescue SUSY, it is both interesting and well motivated to study
the possibility of novel SUSY breaking mechanisms that keep the
Higgs mass protected. In this paper, motivated by the AdS/CFT
correspondence, we propose a mechanism to preserve light scalar
fields using a special form of explicit SUSY breaking.  In fact, we
show that there are moduli -- scalar fields with exactly flat
potentials -- in a non-supersymmetric theory. This is surprising,
and we will explain how quantum corrections cancel for these moduli.

Our model is a three-dimensional supergravity theory in anti-de Sitter (AdS)
spacetime, which is dual to a two-dimensional conformal field theory
(CFT) on the asymptotic boundary.  It incorporates a boundary
deformation, so that  the full action is
\be\la{action} S = S_0 + \frac{h}{2} \int_{\rm bdy} A \wg \tA \,.
\ee
$S_0$ is the action of the undeformed theory in the $AdS_3$ bulk, 
$h$ is the coupling constant governing the strength of the deformation, and $A_\m$, $\tA_\m$ are Chern-Simons gauge fields that respectively satisfy  self-dual and anti-self-dual boundary conditions in the undeformed theory.

The undeformed supergravity theory has at least $\cn=2$
supersymmetry and a gauged $U(1)_L\times U(1)_R$ R-symmetry group.
If we choose $A_\m$ and $\td A_\m$ to be the vector potentials
coupled to the R-symmetry currents, the deformation \er{action}
explicitly breaks all supersymmetries present in the undeformed
theory.  As a physical demonstration of the supersymmetry breaking
effect, we calculate the mass shifts of fields in a
supermultiplet due to \er{action} and show that they are
incompatible with a supersymmetric spectrum. Bulk coupling constants
also shift.

Although supersymmetry is completely broken by the deformation
\er{action}, scalar fields that are moduli in the undeformed theory
continue to have exactly flat potentials after the deformation.  In
particular, this means that these scalar fields remain exactly
massless even after all supersymmetries are broken in the theory.

The key to our mechanism is the Chern-Simons field which has no bulk
degrees of freedom. In AdS/CFT this has the immediate consequence that the
bulk-to-boundary propagator is a ``pure gauge'' $K_{\mu i}(x,\vec w) =
\partial_\mu \Lambda_i(x,\vec w)$ where $x^\mu$ and $w^i$ indicate  bulk
and boundary points, respectively. In Witten diagrams that encode correlation functions,  the bulk derivative
may be integrated by parts. Gauge invariance then ensures that insertions
of $A_\mu$ on a charged line within the bulk cancel among diagrams leaving
boundary contributions  for external charged lines  and no contributions
for external moduli.
We illustrate this by explicit calculation of several relatively simple diagrams in which the final
expression agrees with the OPE calculation in the dual CFT.  We argue that
the same mechanism works for all Witten diagrams.

The existence and number of moduli in the undeformed theory is
determined by its action $S_0$ in $AdS_3$.  A natural way to obtain
such a theory is through string compactifications such as $AdS_3
\times S^3 \times T^4$ \cite{Giveon:1998ns, de Boer:1998pp, Kutasov:1999xu}.  These compactifications
naturally produce moduli; in the $AdS_3 \times S^3 \times T^4$
model, they could be toroidal fluctuations in $T^4$.

Alternatively, one can define the undeformed theory in $AdS_3$ by
its dual CFT.  The two-dimensional CFT has at least $(2,2)$
supersymmetry, and the bulk deformation \er{action} is dual to the
CFT deformation
\be \label{cftdef}
S_{\CFT} = S_{\CFT,0} + \frac{h}{2} \int J \wg
\tJ \,,
\ee
where $J_i$ and $\td J_i$ are the left- and right-moving
R-symmetry currents in the CFT.  This double trace deformation is
exactly marginal \cite{Chaudhuri:1988qb}, so the deformed theory
remains conformal for arbitrary $h$. The deformation also breaks
SUSY.

A particular model of this type has previously been constructed by
taking the near horizon limit of a stack of fundamental strings and
NS5-branes, resulting in an $AdS_3 \times S^3 \times T^4$ solution
with NS fluxes \cite{Giveon:1998ns, de Boer:1998pp, Kutasov:1999xu}.  The deformed theory has motivated
the development of a non-local version of string theory \cite{Aharony:2001pa}, which is
then used to analyze the absence of quantum corrections to the
moduli potential \cite{Aharony:2001dp}. The deformation  \er{action} and its dual \er{cftdef} were  introduced in this context.

One of the main goals of this paper is to provide a bulk field
theory argument to explain why moduli in the undeformed theory
continue to have flat potentials after the deformation.  This allows
us to generalize the particular model of \cite{Aharony:2001dp}
(which has a well-defined string perturbation theory) to virtually
any consistent bulk theory that is a deformation \er{action} of an
$\cn=2$ supergravity theory (with gauged R-symmetry currents and at
least one modulus).  We also explain how the bulk field theory
argument agrees with OPE calculations in the boundary CFT.

%

\section{Basics of our model}

Supergravity models in AdS$_3$ with Chern-Simons dynamics for their vector gauge fields  were first constructed by Achucarro and Townsend in 1986 \cite{Achucarro:1987vz}.
The $\cn =4$ model with R-symmetry group $SU(2)\times SU(2)$ is frequently discussed in the literature \cite{Giveon:1998ns, de Boer:1998pp, Kutasov:1999xu, Aharony:2001pa},  but our model requires only a $U(1)\times \tilde U(1)$ subgroup with gauge fields $A_\m$ and $\tilde A_\m$.
We focus  on  terms in the undeformed action which play a direct role in our calculations, beginning with the Euclidean Chern-Simons action

\bea\label{csaction}
S &=& \frac{1}{8\pi} \int_{\text{bulk}} \[k A\wedge dA - \tilde k\tilde A\wedge d\tilde A\]\,-\, \frac{i}{16\pi}\int_{\text{bdy}}\[k A\wedge *A+ \tilde k\tilde A\wedge *\tilde A\]\\
&=& \nonumber \frac{1}{8\pi} \int_{\text{bulk}} d^3x\,
\e^{\m\r\n}\[k A_\m \pa_\r A_\n- \tilde k (A \leftrightarrow \tilde A)\]  -
\frac{i}{16\pi}\int_{\text{bdy}} d^2w \[k(A_1^2 + A_2^2)+\tilde k (A
\leftrightarrow \tilde A)\]\,. \eea

For integer levels $k,~\tilde k$. the normalization  is that of the $SU(2)$ theory (see \cite{Witten:1988hf}).
As discussed in \cite{Maldacena:2001ky,Aharony:2001dp}, the purpose
of the boundary action is to enforce the condition that the
anti-holomorphic component of $A$ and the holomorphic component of
$\tilde A$ vanish on the boundary.

The bulk theory also contains massive matter multiplets in which the scalar and spinor carry $U(1)\times \td U(1)$ R-charges  $(q,\tilde q)$ and $(q-1, \tilde q)$ or $(q, \tilde q-1)$, respectively.  Charged fields are minimally coupled to $A_\m,~\tilde A_\m$ by covariant derivatives, e.g.  $D_\m \phi = (\pa_\m + iq A_\m+i\tilde q \tilde A_\m )\phi$.

In the introduction we stated that the deformation
\be \label{bulkdef} \cl_{\text{def}} = h A_w \tilde A_{\bar
w} \ee explicitly breaks supersymmetry and is exactly marginal. Both
properties are most simply demonstrated via the dual deformation in
the CFT, namely \be S_{\text{CFT-def}} = h \int d^2w J(w) \tilde
J(\bar w) \,, \ee in which the holomorphic $U(1)$ and anti-holomorphic
$\tilde U(1)$ R-currents appear.  The R-currents are the lowest
components of supermultiplets as is their product. It is well known
that the spacetime integral of the lowest component of a
supermultiplet is not supersymmetric.  The deformation satisfies the
necessary and sufficient conditions for exact marginality
established in \cite{Chaudhuri:1988qb}.  We discuss this in more
detail in Sec.~\re{se:mg}, where we also present calculations within the AdS theory of the $\co(h)$ and $\co(h^2)$ contributions to the two
point function $\< (A_x \tilde A_{\bar x})(x_0,\vec{x})(A_y \tilde
A_{\bar y})(y_0,\vec{y})\>$  as the two points approach the
boundary. The order $h$ correction vanishes by charge conjugation as do all odd orders $h^{2n+1}$. The order $h^2$ amplitude has
divergences in disconnected diagrams only. They are cancelled either
by the vacuum diagrams or by counterterms for the 1-point function. This situation persists to all orders in $h$.

\section{Bulk calculations for the mass correction}
In our two-dimensional CFT, the double trace deformation explicitly breaks
SUSY. The SUSY breaking effect on which we focus  is that  the
conformal dimensions of boson and fermion
operators in the same supermultiplet shift differently due to the deformation. However, to all orders in $h$ there is no such shift for scalar fields that are moduli in the undeformed theory and carry no $R$-charge. Instead the conformal dimensions of their superpartners (modulini) are shifted.

In this section, we carry out explicit calculations in $AdS_3$ and
study perturbative effects due to the explicit SUSY breaking boundary term  (\ref{bulkdef}).
In Witten diagrams this deformation determines an insertion of two
bulk-to-boundary gauge field propagators, for $A$ and $\tilde{A}$
respectively, at one point on the $AdS_3$ boundary.  The propagators are derived in Appendix \re{ap:prop}.

We first calculate the leading order correction to the conformal dimension of a charged scalar which translates to a mass correction of the dual bulk field, at tree level in the bulk couplings. The result matches the
    CFT calculation in \cite{Aharony:2001dp}.
We then undertake a detailed
calculation of the leading order deformation for a modulus field
at the 1-loop level. We  show how the sum of several diagrams
cancels  and leaves the modulus mass untouched. Note that it is equivalent to speak of the conformal dimension of a CFT operator and the mass of the dual bulk field because they are related by the usual AdS/CFT formula (for a scalar in $D=3$ or $d=2$):
\begin{equation}
\Delta = 1 + \sqrt{1 +m^2L^2}\,.
\end{equation}

\subsection{Mass correction of a charged particle}
\label{se:ChargedTree}

\begin{figure}[h]
\begin{center}
\includegraphics[width=0.2\textwidth]{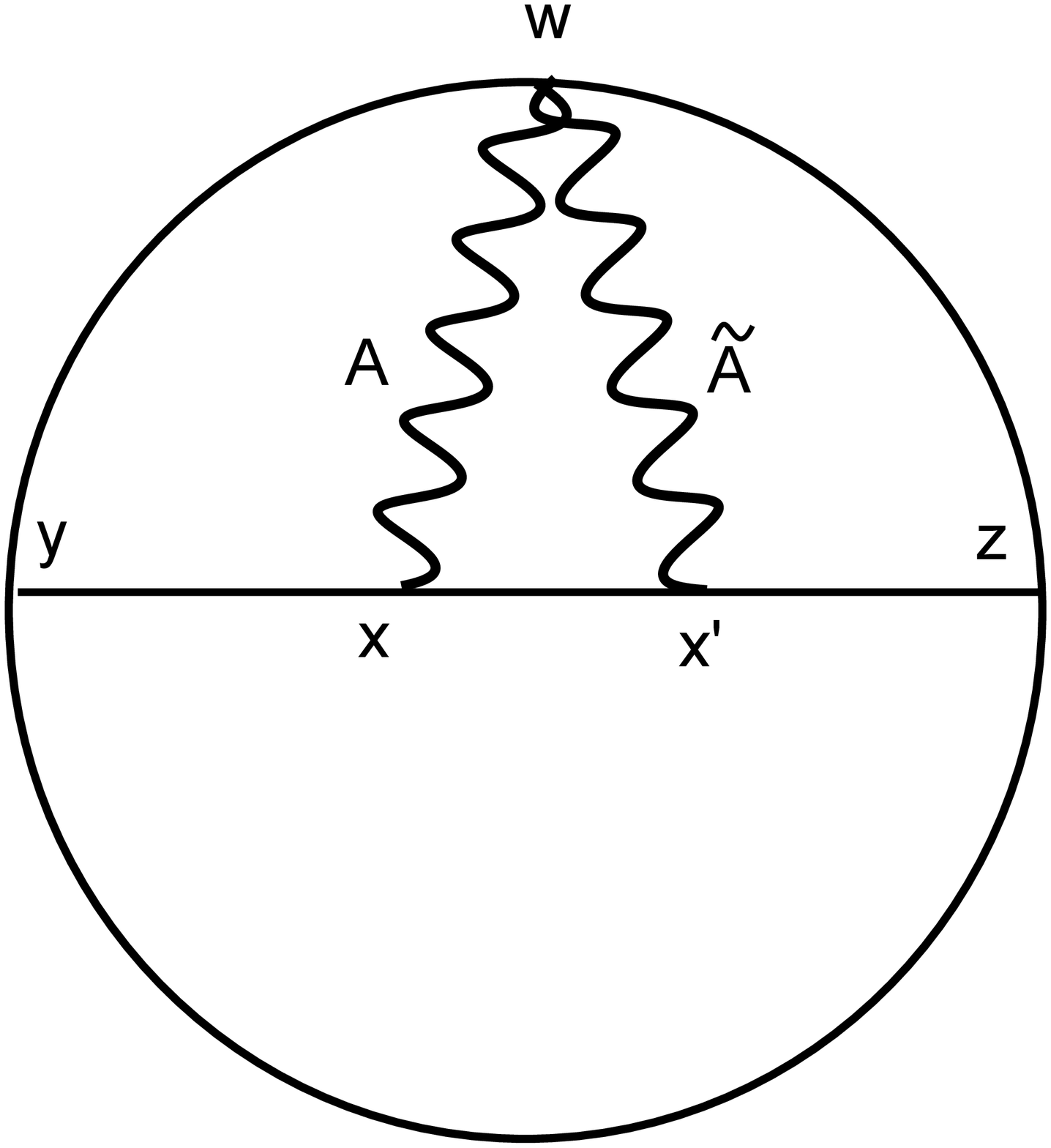}
\hspace*{0.35cm}
\includegraphics[width=0.2\textwidth]{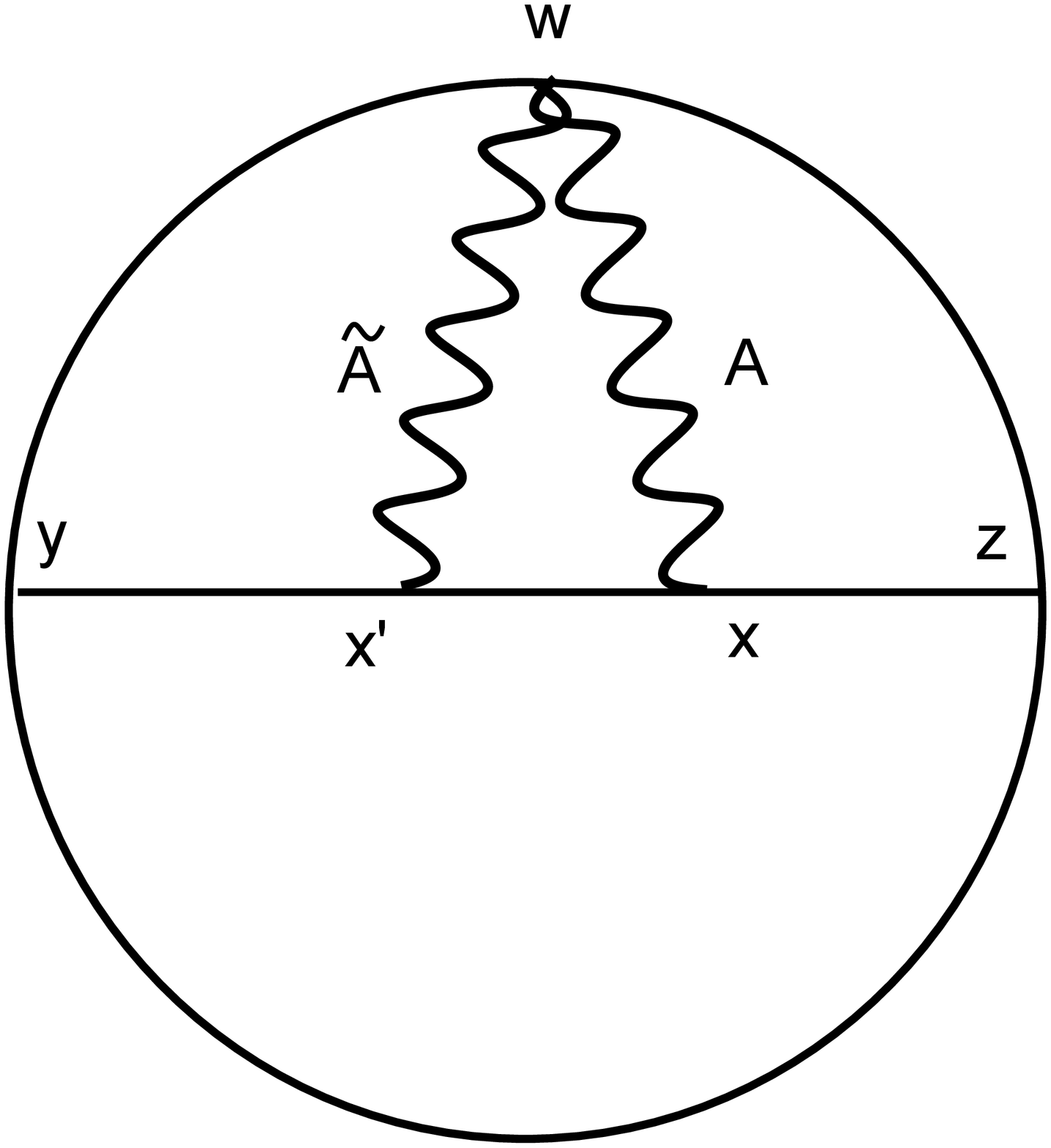}
\hspace*{0.35cm}
\includegraphics[width=0.2\textwidth]{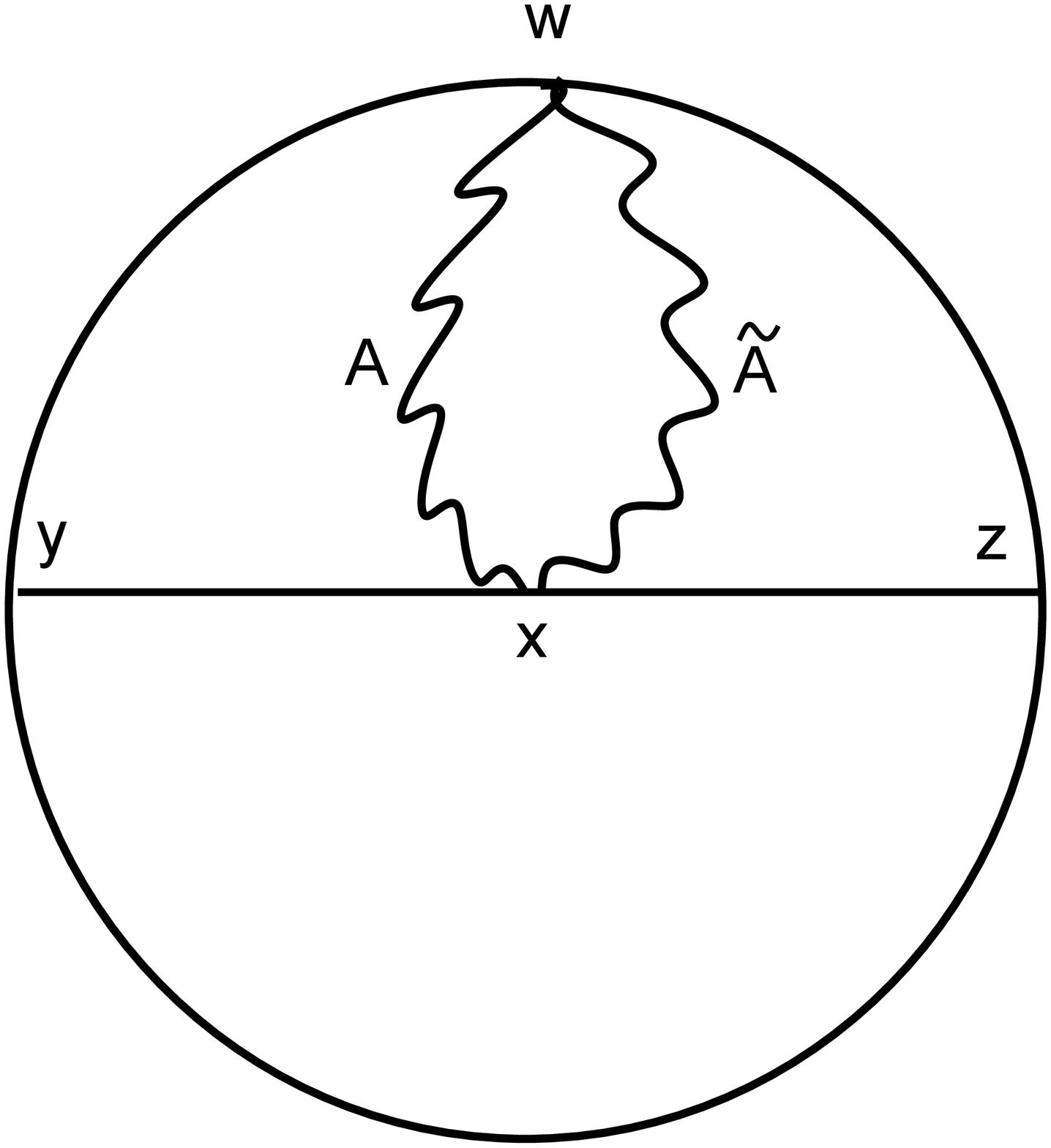}
\caption{The relevant diagrams for the leading order mass deformation
of a charged scalar field.}
\label{fig:Charged}
\end{center}
\end{figure}

In this subsection, we calculate carefully the leading order
correction to the conformal dimension of the CFT operator dual to a charged scalar field in the $AdS_3$ bulk.
The relevant Witten
diagrams are shown in Fig.~\ref{fig:Charged}.
The scalar field is assumed to carry R-charges $(q,\tilde q)$.

The leading order correction to the two-point correlation function
from the first two diagrams is given by the first expression below and then partially integrated using the pure gauge structure of the bulk-to-boundary propagator $K_{\mu,w}(x,\vec{w})= \pa_\mu\Lambda(x,\vec w)$:
\begin{eqnarray}
  \label{eq:ChargeMass}
\delta_h\langle O^\dag_c(\vec{y}) O_c(\vec{z})\rangle &=& h q
\tilde{q}\int d^2w d^3x d^3x'\sqrt{g(x)}\sqrt{g(x')}
K_{\mu,w}(x,\vec{w})\tilde{K}_{\nu,\bar{w}}(x',\vec{w}) \times\nonumber\\
&&\qqu\qqu \times[K_{\Delta}(x,\vec{y})\overset{\leftrightarrow}{\partial^{\mu}}(G_\Delta(x,x')\overset{\leftrightarrow}{\partial'^{\nu}}K_{\Delta}(x',\vec{z}))]+ (\vec{y}\leftrightarrow\vec{z})\nonumber\\
&=& -h q \tilde{q}\int d^2w \frac{d^3x'}{x'_0}d^3x\sqrt{g(x)}
\Lambda_{w}(x,\vec{w})\tilde{K}_{\nu,\bar{w}}(x',\vec{w}) \times\nn\\
&&\qqu\qqu \times[K_{\Delta}(x,\vec{y})\overset{\leftrightarrow}{\square}(G_\Delta(x,x')\overset{\leftrightarrow}{\partial'_\nu}K_{\Delta}(x',\vec{z}))]
+ (\vec{y}\leftrightarrow\vec{z}) \nn\\
&& -h q \tilde{q}\int d^2w\frac{d^3x'}{x'_0} \lim_{x_0\to
0}\frac{d^2x}{x_0}
\Lambda_{w}(x,\vec{w})\tilde{K}_{\nu,\bar{w}}(x',\vec{w}) \times\nn\\
&&\qqu\qqu \times[K_{\Delta}(x,\vec{y})\overset{\leftrightarrow}\partial_{0}(G_\Delta(x,x')\overset{\leftrightarrow}{\partial'_\nu}K_{\Delta}(x',\vec{z}))]
+ (\vec{y}\leftrightarrow\vec{z}) \,.
\end{eqnarray}
Here $K_{\Delta}$ and $G_{\Delta}$ are the bulk-to-boundary and bulk
propagators of a scalar field, for which we will need only the form \er{btbdy}.

From Eq.~(\ref{eq:ChargeMass}) we see that the corrections to the
two-point correlation function of a charged scalar field break into
two parts: one is the bulk contribution after partial integration,
and the other is the contribution from the boundary. Let us first focus on
the bulk part:
\begin{eqnarray}
  \label{eq:ChargeMassBulk}
&&\delta_h\langle O^\dag_c(\vec{y}) O_c(\vec{z})\rangle_{bulk}\nn\\
&=& h q \tilde{q}\int d^2w \frac{d^3x'}{x'_0}d^3x\sqrt{g(x)}
\Lambda_{w}(x,\vec{w})\tilde{K}_{\nu,\bar{w}}(x',\vec{w}) \times \nonumber\\
&&\qqu\qqu \times[K_{\Delta}(x,\vec{y})(\delta^{3}(x,x')\overset{\leftrightarrow}{\partial'_\nu}K_{\Delta}(x',\vec{z}))]+ (\vec{y}\leftrightarrow\vec{z})\nonumber\\
&=& h q \tilde{q}\int
d^2w\frac{d^3x'}{x'_0}\Lambda_{w}(x',\vec{w})
[K_{\Delta}(x',\vec{y})\overset{\leftrightarrow}{\partial'_\nu}K_{\Delta}(x',\vec{z})]\tilde{K}_{\nu,\bar{w}}(x',\vec{w})+ (\vec{y}\leftrightarrow\vec{z})  \nn\\
&&-h q \tilde{q}\int
d^2w\frac{d^3x'}{x'_0}[{\partial'_\nu}\Lambda_{w}(x',\vec{w})]
K_{\Delta}(x',\vec{y})K_{\Delta}(x',\vec{z})\tilde{K}_{\nu,\bar{w}}(x',\vec{w})+ (\vec{y}\leftrightarrow\vec{z})\nonumber\\
&=&-2h q \tilde{q}\int d^2w d^3x'
\sqrt{g(x')}g^{\rho\nu}(x')K_{\rho,w}(x',\vec{w})
K_{\Delta}(x',\vec{y})K_{\Delta}(x',\vec{z})\tilde{K}_{\nu,\bar{w}}(x',\vec{w})\,.\nonumber\\
\end{eqnarray}
In the first step of the calculation, we used the following
properties of scalar bulk and bulk-to-boundary propagators:
\begin{eqnarray}  \label{green}
  \label{eq:propagator}
(\square-m^2) G_{\Delta}(x,x')&=&-\delta^3(x,x')/\sqrt{g}\nonumber\\
(\square-m^2) K_{\Delta}(x,\vec{y})&=&0\,.
\end{eqnarray}
We then find that the bulk part of the correction
cancels precisely with the contribution from the seagull diagram in Fig.~\re{fig:Charged}. Thus
the only correction to the 2-point correlation function comes
from the boundary terms:
\begin{eqnarray}
  \label{eq:ChargeMassBdy}
\delta_h\langle O^\dag_c(\vec{y}) O_c(\vec{z})\rangle_{bdy} &=&-h q
\tilde{q}\int d^2w\frac{d^3x'}{x'_0} \lim_{x_0\to 0}\frac{d^2x}{x_0}
\Lambda_{w}(x,\vec{w})\tilde{K}_{\nu,\bar{w}}(x',\vec{w}) \times\nonumber\\
&&\qqu \times
[K_{\Delta}(x,\vec{y})\overset{\leftrightarrow}\partial_{0}(G_\Delta(x,x')\overset{\leftrightarrow}{\partial'_{\nu}}K_{\Delta}(x',\vec{z}))]+
(\vec{y}\leftrightarrow\vec{z})\,.
\end{eqnarray}
To proceed with the calculation, the following equations
 are useful:
\begin{eqnarray}
  \label{eq:usefulEq}
\lim_{x_0\to 0}\ x_0^{\Delta-d}K_{\Delta}(x,\vec{y})&=&\delta^2(\vec{x},\vec{y}) \,,\nonumber\\
\lim_{x_0'\to 0}\ (2\Delta-d) x_0'^{-\Delta}
 G_{\Delta}(x,x')&=&K_{\Delta}(x,\vec{x}') \,,\nonumber\\
\partial_{x_0}K_{\Delta}(x,\vec{y})&=
&\frac{\Delta}{x_0}K_{\Delta}(x,\vec{y})-2\Delta\frac{C_\Delta}{C_{\Delta+1}}K_{\Delta+1}(x,\vec{y}) \,,
\end{eqnarray}
where the explicit form of $C_\D$ is given in \er{btbdy}.
Then  (\ref{eq:ChargeMassBdy}) can be written as
\begin{eqnarray}
  \label{eq:ChargeMassBdy2}
&&\delta_h\langle O^\dag_c(\vec{y}) O_c(\vec{z})\rangle_{bdy} \nn\\
&=&\[-\frac{\Delta}{2(\Delta-1)}+\(\Delta-\frac{2\Delta
C_\Delta}{C_{\Delta+1}}\)\frac{1}{2(\Delta-1)}\]\ h q \tilde{q}\times \nonumber\\
&&\qqu\times\int d^2w
\Lambda_{w}(0,\vec{y},\vec{w})\int\frac{d^3x'}{x'_0}
[K_\Delta(x',\vec{y})\overset{\leftrightarrow}{\partial'_{\nu}}K_{\Delta}(x',\vec{z})]\tilde{K}_{\nu,\bar{w}}(x',\vec{w}) + (\vec{y}\leftrightarrow\vec{z})\nonumber\\
&=&-h q \tilde{q} \int d^2w
\Lambda_{w}(0,\vec{y},\vec{w})\int\frac{d^3x'}{x'_0}
[K_\Delta(x',\vec{y})\overset{\leftrightarrow}{\partial'_{\nu}}K_{\Delta}(x',\vec{z})]\partial'_\nu \tilde{\Lambda}_{\bar{w}}(x',\vec{w}) + (\vec{y}\leftrightarrow\vec{z})\nonumber\\
&=&h q \tilde{q} \int d^2w \Lambda_{w}(0,\vec{y},\vec{w})\int
d^3x'\sqrt{g(x')}
[K_\Delta(x',\vec{y})\overset{\leftrightarrow}{\square'}K_{\Delta}(x',\vec{z})]\tilde{\Lambda}_{\bar{w}}(x',\vec{w})+ (\vec{y}\leftrightarrow\vec{z}) \nn\\
&&\qu+h q \tilde{q} \int d^2w
\Lambda_{w}(0,\vec{y},\vec{w})\lim_{x'_0\to 0}\int
\frac{d^2x'}{x'_0}
[K_\Delta(x',\vec{y})\overset{\leftrightarrow}{\partial'_{0}}K_{\Delta}(x',\vec{z})]\tilde{\Lambda}_{\bar{w}}(x',\vec{w})+
(\vec{y}\leftrightarrow\vec{z})\,.\nonumber\\
\end{eqnarray}
Note that the first line in the final equality vanishes
due to Eq.~(\ref{eq:propagator}). Again only boundary contributions
survive. Applying the last equation in Eq.~(\ref{eq:usefulEq}), one
finds
\begin{eqnarray}
  \label{eq:ChargeMassBdy3}
&&\delta_h\langle O^\dag_c(\vec{y}) O_c(\vec{z})\rangle_{bdy} \nn\\
&=& h q \tilde{q} \int d^2w
\Lambda_{w}(0,\vec{y},\vec{w})\lim_{x'_0\to 0}\int
\frac{d^2x'}{x'_0}\tilde{\Lambda}_{\bar{w}}(x',\vec{w}) \times\nonumber\\
&&\qu\times\[-2\Delta\frac{C_\Delta}{C_{\Delta+1}}K_{\Delta}(x',\vec{y})K_{\Delta+1}(x',\vec{z})+2\Delta\frac{C_\Delta}{C_{\Delta+1}}K_{\Delta+1}(x',\vec{y})K_{\Delta}(x',\vec{z})\] +(\vec{y}\leftrightarrow\vec{z})\nonumber\\
&=& h q \tilde{q} \int d^2w \Lambda_{w}(0,\vec{y},\vec{w})\int
d^2x'\tilde{\Lambda}_{\bar{w}}(0,\vec{x}',\vec{w})\, 2\Delta\frac{C_\Delta^2}{C_{\Delta+1}} \[\frac{\delta^{2}(\vec{x}',\vec{y})}{|\vec{x}'-\vec{z}|^{2\Delta}}-\frac{\delta^{2}(\vec{x}',\vec{z})}{|\vec{x}'-\vec{y}|^{2\Delta}}\] +(\vec{y}\leftrightarrow\vec{z})\nonumber\\
&=& \frac{2(\Delta-1)^2}{\pi} \frac{h q
\tilde{q}}{|\vec{y}-\vec{z}|^{2\Delta}}\int d^2w
\[\frac{1}{|y-w|^2}-\frac{1}{(y-w)}\frac{1}{(\bar{z}-\bar{w})}\]+(\vec{y}\leftrightarrow\vec{z})\nonumber\\
&=&\(2\pi h q \tilde{q}\
\textrm{log}\frac{|y-z|^2}{|a|^2}\)\frac{2(\Delta-1)^2}{\pi}\frac{1}{|\vec{y}-\vec{z}|^{2\Delta}}\nonumber\\
&=& \(2\pi h q \tilde{q}\,\textrm{log}\frac{|y-z|^2}{|a|^2}\)\langle
O^\dag_c(\vec{y}) O_c(\vec{z})\rangle_0 \,,
\end{eqnarray}
where $a$ is the short-distance regulator for the integral. There is a subtlety  in the
 boundary limit of  the product $K_\Delta K_{\Delta+1}$. One can either take the $\delta$ function
limit of $K_\Delta$ and study the boundary limit of $K_{\Delta+1}$ or vice versa, depending on the position of
$\vec{x}'$ when we take the limit $x_0'\rightarrow 0$.  However, the first choice
 vanishes in the limit. Thus only the latter choice
contributes and gives the fourth line of Eq.~(\ref{eq:ChargeMassBdy3}).

Note that the two-point function of the undeformed theory appears as a factor.
From the coefficient of the logarithm, one can identify the shift in
$\Delta$
\begin{equation}
\label{eq:masschange}
\delta_h\Delta = -2\pi h q \tilde q \,.
\end{equation}
This result agrees perfectly with the dimension shift obtained in \cite{Aharony:2001dp}.

At this point we can see without repeating the calculation that the
leading correction to the 2-point function of the spinor
superpartner $\Psi_c$ of $O_c$ must be
\begin{equation}\label{spinor}
\delta_h\langle\Psi_c(\vec y) \bar\Psi_c(\vec z)\rangle= \(2\pi h (q-1)
(\tilde{q})\,\textrm{log}\frac{|y-z|^2}{|a|^2}\)\langle\Psi_c(\vec
y)\bar\Psi_c(\vec z)\rangle_0\,.
\end{equation}
The last factor is the undeformed spinor two-point function. To
justify this claim we note that the calculation proceeds by the same
steps of partial integration and use of the Ward identity. The spinor case is even simpler than the scalar case because there are
no seagull diagrams and it is not necessary to differentiate
(with $\partial_{x_0}$) the spinor bulk-to-boundary propagator.  The result
(\ref{spinor})  differs from Eq.~(\ref{eq:masschange}) for the
scalar only via the change in the R-charges, i.e. the scalar charges
$(q,\tilde q)$ are replaced by $(q-1, \tilde q)$ for the fermion.

\subsection{Mass correction for moduli fields}\label{se:ModuliMass}
In this subsection, we focus on bulk  moduli fields which are
neutral under $R$-symmetry. We show that the shift $\delta_h\Delta$ of
such a field vanishes at 1-loop order. To simplify the calculation,
we assume that the moduli
 couple to charged scalar particles through the 3-point vertex\footnote{The $U(1)$ Ward identity implies that the result is also valid for derivative vertices such as $\cl' \sim \phi_m D_\m\phi_c^\dagger D^\m\phi_c$, although the diagrammatic analysis is more complicated.}
\begin{equation}
  \label{eq:Yukawa}
\cl\supset y \phi_{m}\phi_c^{\dag}\phi_c
\end{equation}
where $\phi_m$ is an $R$-neutral modulus field and $\phi_c$ has
non-zero R-charges $(q, \tilde{q})$.
As we have shown in the previous section, the
mass of $\phi_c$ is modified by the SUSY breaking deformation
according to Eq.~(\ref{eq:masschange}). One might expect that moduli
masses will also shift due to SUSY breaking effects in loop
diagrams. However we will show that when all contributing diagrams
are included,  SUSY breaking effects cancel and leave the moduli
untouched.

\begin{figure}[h]
\begin{center}
\includegraphics[width=0.2\textwidth]{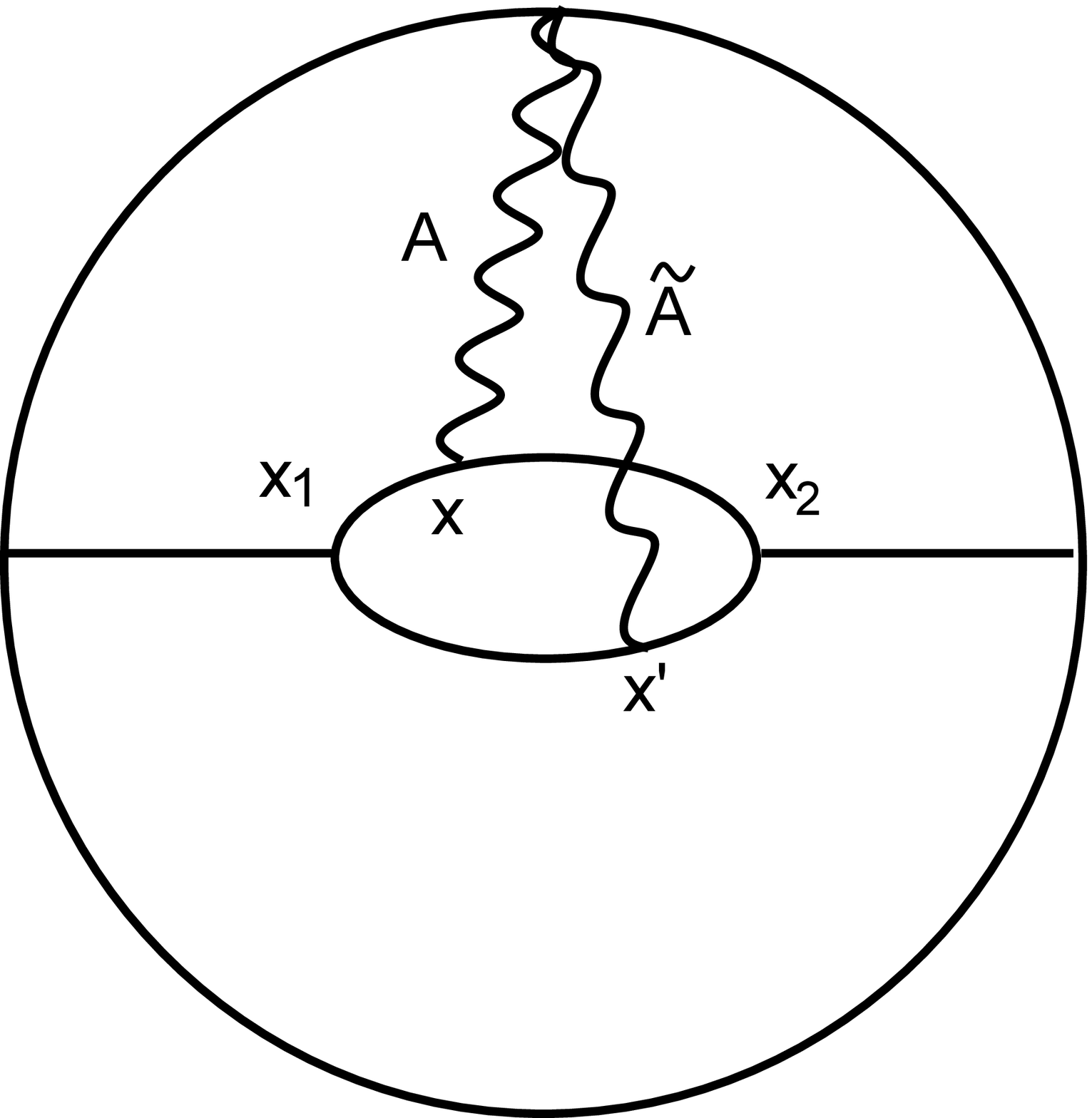}
\hspace*{0.35cm}
\includegraphics[width=0.2\textwidth]{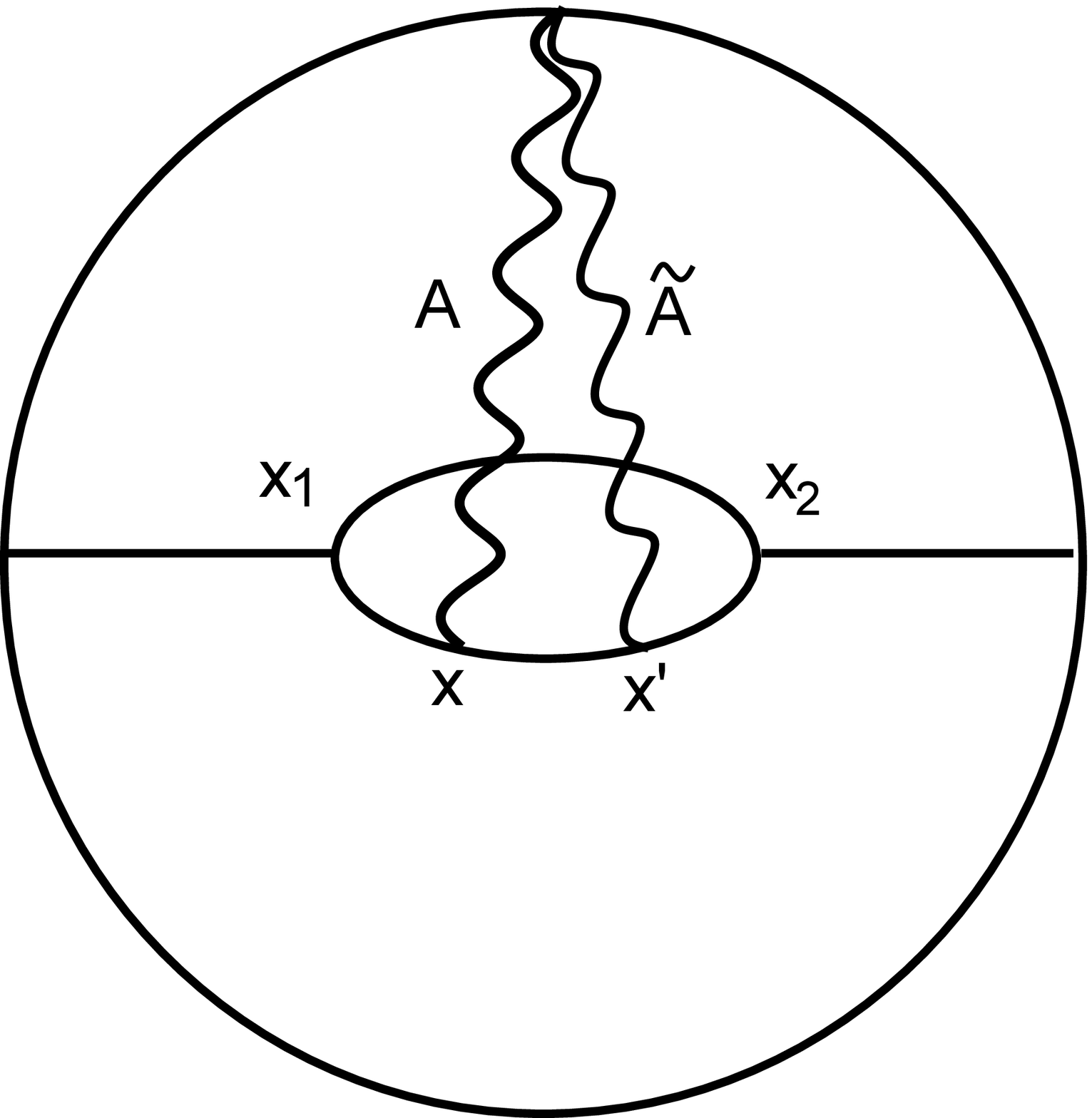}
\hspace*{0.35cm}
\includegraphics[width=0.2\textwidth]{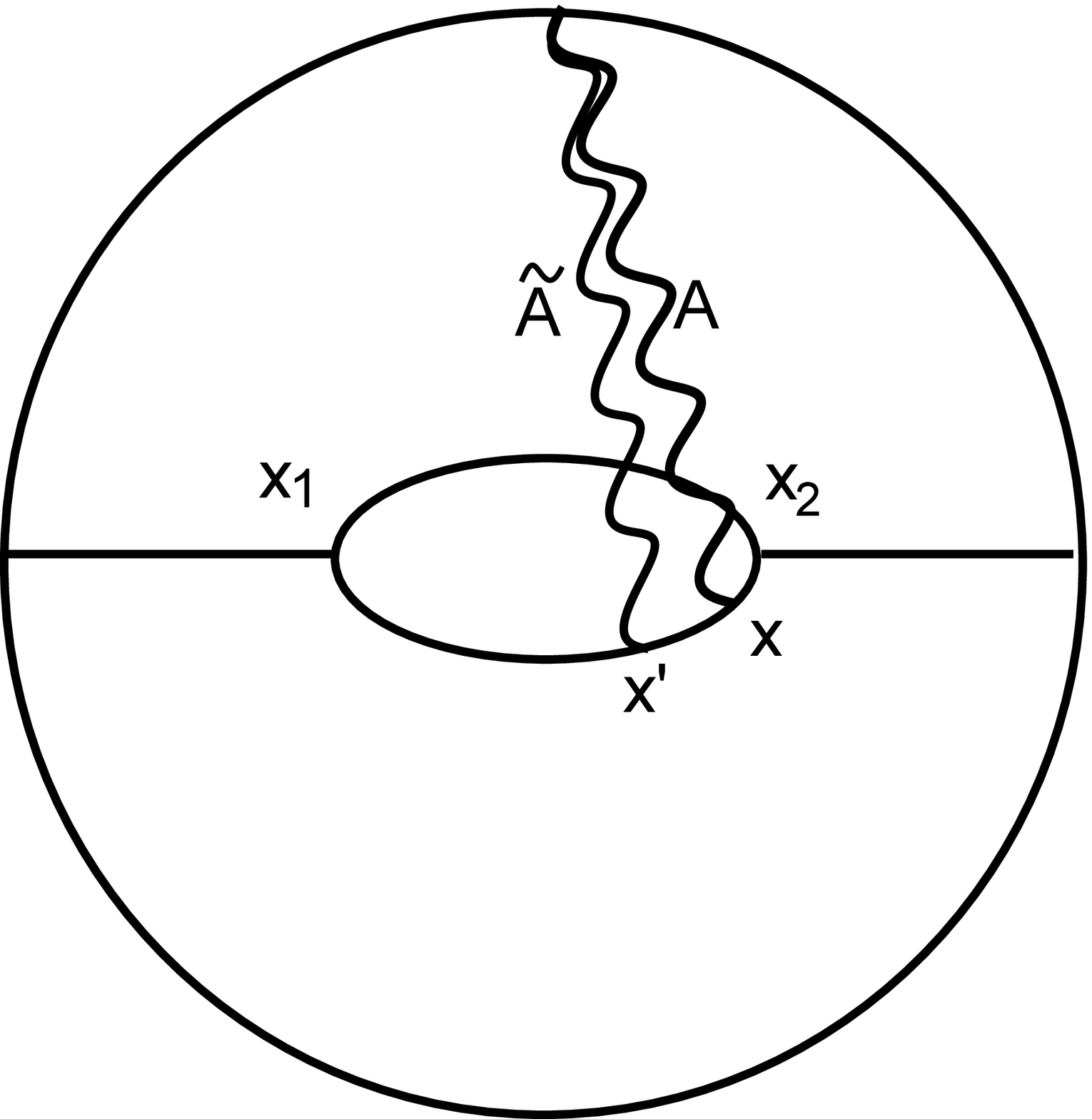}
\hspace*{0.35cm}
\includegraphics[width=0.2\textwidth]{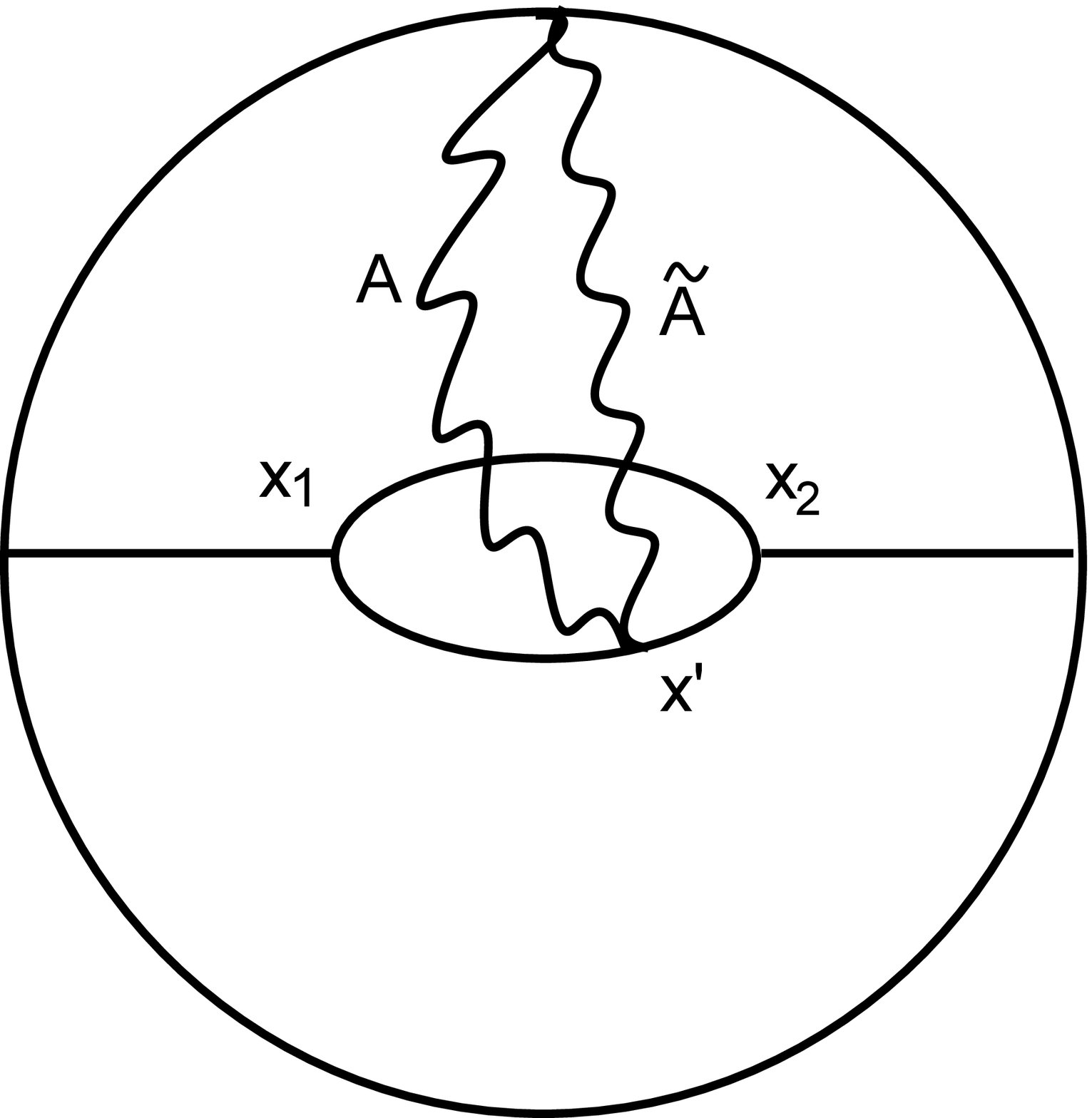}
\caption{The relevant diagrams for calculating the leading order
deformation of the 1-loop self-energy correction. Here we fix the
position of $x'$ while moving $x$ around the loop of the charged
field. 
 \label{fig:LoopLeading}}
\end{center}
\end{figure}

In Fig.~\ref{fig:LoopLeading}, we list the  relevant diagrams. To
exhibit the cancellation,  we fix the position of the $\tilde{A}$
propagator and add the amplitudes for diagrams in which the $A$
propagator is attached at all possible positions on the charged $\phi_c$
loop. Since moduli fields are neutral, $A$ and $\tilde{A}$ cannot
couple to the external lines of Fig.~\ref{fig:LoopLeading}.
The last diagram of the figure is determined by the seagull vertex  $2 q\tilde{q}\sqrt{g}
A_\mu\tilde{A}^\mu\phi_c^\dag\phi_c.$

In our calculation, we focus first on the integration of the
end point position $x$ of $A_\mu$ in each diagram. Thus we temporarily ignore factors  in the amplitude which  do not depend on the bulk 3-vector $x$.   Those factors are denoted by (...).
We start from the
simplest case, i.e. the first diagram in Fig.~\ref{fig:LoopLeading}:

\begin{eqnarray}
  \label{eq:PhiPhi1}
\delta_h\langle O_m O_m\rangle_1 &=& h q \tilde{q}\int d^2w d^3x
d^3x'\sqrt{g(x)}\sqrt{g(x')}
K_{\mu,w}(x,\vec{w})G_\Delta(x_1,x)\overset{\leftrightarrow}{\partial^\mu}
G_\Delta(x,x_2)(...)\nonumber\\
   &=& -h q \tilde{q}\int d^2w d^3x
d^3x'\sqrt{g(x)}\sqrt{g(x')} \Lambda_w(x,\vec{w})
G_\Delta(x_1,x)\overset{\leftrightarrow}{\square}
G_\Delta(x,x_2)(...)  \nn\\
&&\qqu -hq\tilde{q}\int d^2w d^3 x' \sqrt{g(x')}\lim_{x_0\to 0}\frac{d^2
x}{x_0}\Lambda_\omega(0,\vec{x},\vec{w})G_{\Delta}(x_1,x)\overset{\leftrightarrow}{\partial_0}G_{\Delta}(x,x_2)(...)\nonumber\\
   &=& h q \tilde{q}\int d^2w d^3x
d^3x'\sqrt{g(x)}\sqrt{g(x')} \Lambda_w(x,\vec{w}) [G_\Delta(x_1,x)\delta^3(x,x_2)-(x_1\lra x_2)](...)\nonumber\\
   &=&
h q \tilde{q}\int d^2w
d^3x'\sqrt{g(x')}[\Lambda_w(x_2,\vec{w}) -
\Lambda_w(x_1,\vec{w})]G_\Delta(x_1,x_2)(...) \,.
\end{eqnarray}
On the second line, we have used the pure gauge structure
$K_{\mu,w}(x,\vec{w})= \partial_\mu \Lambda_w(x,\vec w)$ and
integrated by parts. If $\Delta \ge 0$, which is satisfied
automatically in a unitary CFT, the boundary term on the third line vanishes due to cancellations in the boundary limit of
$ G_\Delta(x_1,x)\overset{\leftrightarrow}{\partial_0}
G_\Delta(x,x_2)$. On the fourth line, we have used the equation
for $G_\D$ indicated in (\ref{eq:propagator}). We see how the
calculation is greatly simplified due to the pure gauge feature of
$K_{\mu,w}$ and the Ward identity. In the final result the gauge
field insertion is pinned at the end points of the charged
particle's propagators.

Now we move on to calculate more complicated
cases, i.e. the second and third diagrams of Fig.~\ref{fig:LoopLeading}.
\begin{eqnarray}
  \label{eq:PhiPhi2}
\delta_h\langle O_m O_m\rangle_2 &=&- h q \tilde{q}\int d^2w d^3x
d^3x'\sqrt{g(x)}\sqrt{g(x')} \times \nn\\
&&\qqu\qqu\qqu \times K_{\mu,w}(x,\vec{w})G_\Delta(x_1,x)\overset{\leftrightarrow}{\partial^\mu}
[G_\Delta(x,x')\overset{\leftrightarrow}{\partial'^{\nu}}G_\Delta(x',x_2)](...)\nonumber\\
&=&  h q \tilde{q}\int d^2w d^3x d^3x'\sqrt{g(x)}\sqrt{g(x')}\times\nn\\
&&\qqu\qqu\qqu\times
\Lambda_w(x,\vec{w})G_\Delta(x_1,x)\overset{\leftrightarrow}{\square}
[G_\Delta(x,x')\overset{\leftrightarrow}{\partial'^{\nu}}G_\Delta(x',x_2)](...)\nonumber\\
&=& -h q \tilde{q}\int d^2w d^3x'\sqrt{g(x')}
[\Lambda_w(x',\vec{w})G_\Delta(x_1,x')]\overset{\leftrightarrow}{\partial'^{\nu}}G_\Delta(x',x_2)(...)  \nn\\
&& +h q \tilde{q} \int d^2w
d^3x'\sqrt{g(x')}\Lambda_w(x_1,\vec{w})
G_\Delta(x_1,x')\overset{\leftrightarrow}{\partial'^{\nu}}G_\Delta(x',x_2)(...) \,.
\end{eqnarray}
Similar calculations give
\begin{eqnarray}
  \label{eq:PhiPhi3}
\delta_h\langle O_m O_m\rangle_3 &=& -h q \tilde{q}\int d^2w
d^3x'\sqrt{g(x')}
[\Lambda_w(x',\vec{w})G_\Delta(x_2,x')]\overset{\leftrightarrow}{\partial'^{\nu}}G_\Delta(x',x_1)(...) \nn\\
&& +h q \tilde{q} \int d^2w
d^3x'\sqrt{g(x')}\Lambda_w(x_2,\vec{w})
G_\Delta(x_2,x')\overset{\leftrightarrow}{\partial'^{\nu}}G_\Delta(x',x_1)(...) \,.
\end{eqnarray}
After carefully putting back the non-$x$ dependent parts of the
equations, the sum of last terms in Eq.~(\ref{eq:PhiPhi2}) and Eq.
(\ref{eq:PhiPhi3}) precisely cancels Eq.~(\ref{eq:PhiPhi1}). Thus
\begin{eqnarray}
  \label{eq:PhiPhiSum}
\sum_{i=1}^3 \delta_h\langle O_m O_m\rangle_i &=& -h q
\tilde{q}\int d^2w d^3x'\sqrt{g(x')}
[\Lambda_w(x',\vec{w})G_\Delta(x_1,x')]\overset{\leftrightarrow}{\partial'^{\nu}}G_\Delta(x',x_2)(...) \nn\\
&&-h q \tilde{q}\int d^2w d^3x'\sqrt{g(x')}
[\Lambda_w(x',\vec{w})G_\Delta(x_2,x')]\overset{\leftrightarrow}{\partial'^{\nu}}G_\Delta(x',x_1)(...)\nonumber\\
&=&2 h q \tilde{q}\int d^2w d^3x'\sqrt{g(x')}
\partial'^{\nu}\Lambda_w(x',\vec{w})G_\Delta(x_1,x')G_\Delta(x',x_2)(...)\,.
\end{eqnarray}
Interestingly, this is precisely the opposite contribution from
the seagull vertex, i.e. the last diagram in Fig.~\ref{fig:LoopLeading}.
Thus adding up all the contributions, we clearly see the cancellation
of SUSY breaking effects in the mass shift of moduli
fields.

Similar arguments can be applied to fermionic charged particles in
the loop, where the calculation is easier due to the lack of the seagull
vertex. Furthermore, the sum of diagrams for any $n$-point
correlation function of moduli is unaffected by the SUSY breaking term of (\ref{action}).

The modulino partner of a modulus field  carries R-charges $(-1,0)$ or $(0,-1)$.  Since $q\tilde q =0$, its mass shift  vanishes to order $h$, but there are mass corrections of order $h^{2n}$ for all $n$ as we argue in Sec. 5 below.  The modulus mass remains zero to all orders.

\subsection{General structure of the mass correction}

Let us now consider the mass correction of a bulk field with R-charge $(q,\td q)$.  We now argue that the order $h$ correction to any Witten
diagram with $R$-charge conserving bulk vertices has the same
structure as the simple result Eq.~(\ref{eq:ChargeMassBdy3}).  This
structure is
\bea \label{bdypin} \delta_h\langle O^\dag_c (\vec y) O_c(\vec z)\rangle&=&
\langle O^\dag_c (\vec y) O_c(\vec z)\rangle_0 \bigg\{\int d^2w
\Lambda_{w}(0,\vec{y},\vec{w}) \[\tilde{\Lambda}_{\bar{w}}(0,\vec{y},\vec{w})-\tilde{\Lambda}_{\bar{w}}(0,\vec{z},\vec{w})\] \,+\, (\vec y \leftrightarrow \vec z)\bigg\}\nonumber\\
&=&  \langle O^\dag_c (\vec y) O_c(\vec z)\rangle_0 \(2\pi h q
\tilde{q}\,\textrm{log}\frac{|y-z|^2}{|a|^2}\) \,.
\eea
Here $ \langle O^\dag_c (\vec y) O_c(\vec z)\rangle_0$ is the
contribution to the 2-point function from the Witten diagram in the undeformed theory. Thus the shift in the conformal dimension due to the deformation
is again
\begin{equation}
\delta_h\Delta = -2\pi h q \tilde q \,.
\end{equation}

\begin{figure}[h]
\begin{center}
\includegraphics[width=0.2\textwidth]{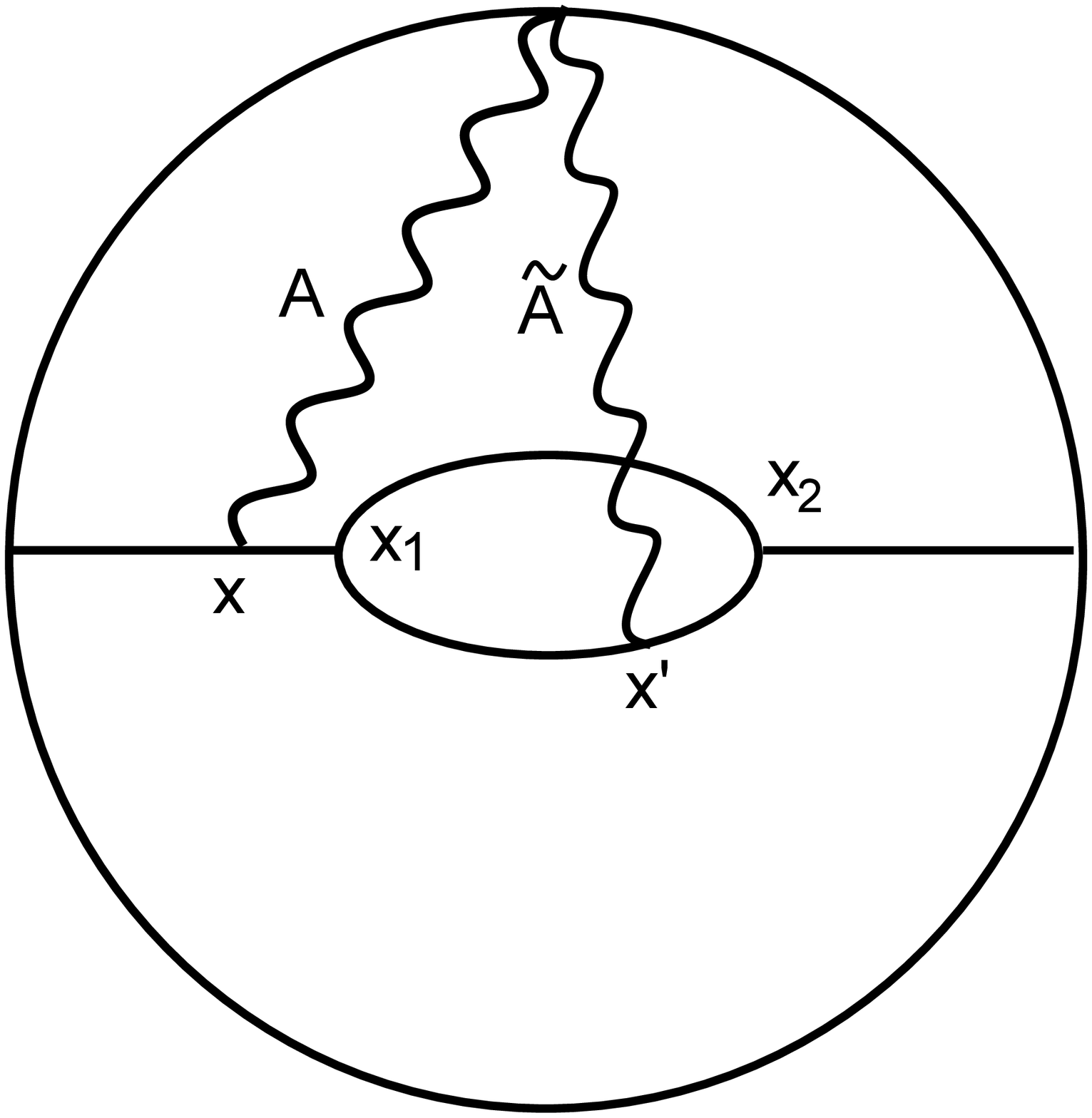}
\hspace*{0.35cm}
\includegraphics[width=0.2\textwidth]{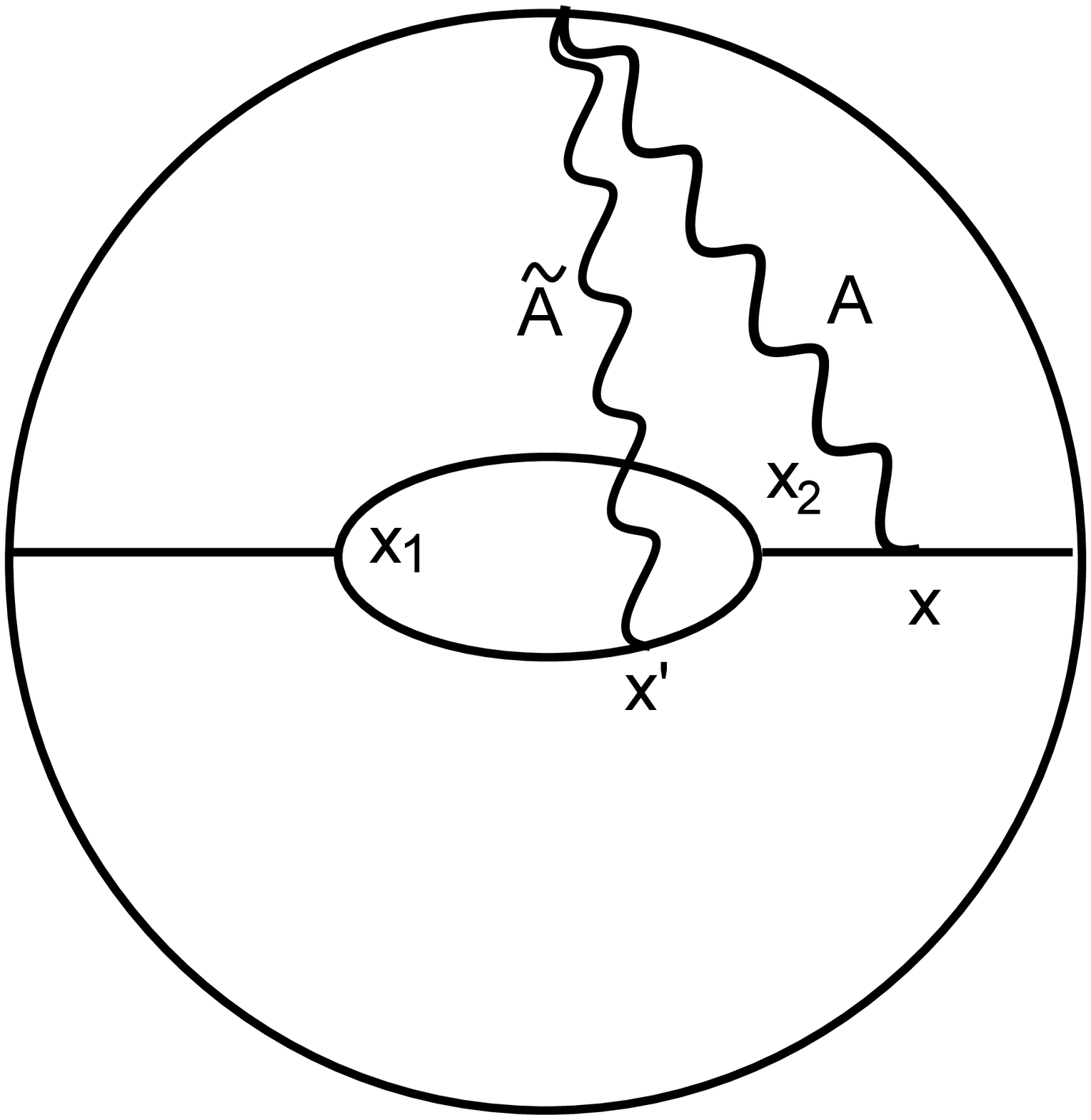}
\hspace*{0.35cm} \caption{Some additional diagrams needed to calculate
 the order $h$ deformation of the 1-loop self-energy correction for a scalar with R-charge $(q,\tilde q)$. 
Diagrams with seagull vertices on external
lines are also needed.} \label{fig:ChargeGeneral}
\end{center}
\end{figure}


The essential principles  of the argument are: 1) the pure gauge
structure of the bulk-to-boundary propagator $K_{\mu,w} (x, \vec w)
=\partial_\mu \Lambda_w(x,\vec w)$, 2) partial integration of
$\partial_\mu$ and the Ward identity,  3) the use of (\ref{green}),
and 4) $R$-charge conservation at each vertex. These principles work
quite generally, but it is useful to visualize it in the following specific example.
Let us choose a bulk theory with a cubic coupling of three
charged scalars ${\cal L}_{\rm cubic} \sim \phi_{(q,\tilde q)}\phi_{(q_1,\tilde
q_1)}\phi_{(q_2,\tilde q_2)} + h.c.$ with $q+q_1+q_2 = \tilde q
+\tilde q_1+\tilde q_2 =0$. We work with the order $h$ deformation
of the 1-loop self-energy diagram for the field $\phi_{(q,\tilde
q)}$.  The diagrams that we need are those of Fig.~\re{fig:LoopLeading} combined with those of Fig.~\re{fig:ChargeGeneral} in
which one or two gauge bosons are coupled to the external lines.

We first consider the subset of diagrams in which the $\tilde A_\nu$
vertex is fixed, and $A_\mu$ is attached at all possible positions.
We have already seen in Sec.~\ref{se:ChargedTree} and
\ref{se:ModuliMass} how principles 1)-3) operate. When applied at a
given insertion point of $A_\mu(x)$ they allow us to integrate over
the bulk position $x$. The result is a sum of two terms in which the
factor $\Lambda_w(x,\vec w)$ is pinned either at the adjacent bulk vertices
if the insertion is on an internal line, or at the boundary and the
adjacent vertex if the insertion is on an external line.  After
applying this procedure to all insertion points of $A_\mu$,  one finds that each
bulk vertex acquires the numerical factor $q+q_1+q_2$ which
vanishes!   Therefore, only diagrams where $\Lambda_w$ is pinned at the boundary points of the
two external lines survive. There remains a smaller set of diagrams
in which $\tilde A_\nu$ is inserted at all possible positions.  When
the procedure 1)-3) is applied to these, one is left with the boundary
factors in the $\vec w$ integral in (\ref{bdypin}) times the value
of the undeformed diagram.  It is clear that this argument applies to all loop orders in the bulk.  Furthermore, we may generalize the calculation to higher orders in $h$ by repeating this procedure.


\section{Boundary CFT calculation for the conformal dimension}

In this section, we use the operator product expansion (OPE) to
calculate the shift of the conformal dimension of operators in the CFT.
We  show that such shifts are  induced by the SUSY breaking
deformation
\begin{eqnarray}
  \label{eq:CFTperturb}
\delta S_{CFT}=h\int d^2z J(z)\tilde{J}(\bar{z}) \,.
\end{eqnarray}
This deformation involves the currents of the $U(1)\times \td U(1)$ R-symmetry group.
Conformal dimensions of operators that are charged under both $U(1)$'s receive a leading
order correction in $h$. If an operator is charged only under one of the
$U(1)$s, its conformal dimension is modified at the next order $h^2$.

%

Many effects of the deformation can
be calculated exactly because $J(x)$ and $\tilde{J}(\bar{x})$ can be
bosonized, i.e.\footnote{Note that our normalization of the current $J$ (and $\tilde J$) is consistent with it being a component of an $SU(2)$ current, and may differ from conventions used elsewhere in the literature such as \cite{Aharony:2001dp}.}
\begin{eqnarray}
  \label{eq:bosonize}
J(z)&=& i\sqrt{k}\partial_z\eta(z) \,,\nonumber\\
\tilde{J}(\bar{z})&=& i\sqrt{\tilde k}\partial_{\bar{z}}\tilde{\eta}(\bar{z}) \,,
\end{eqnarray}
where $\eta$ and $\tilde{\eta}$ are canonically normalized scalar
fields with the OPEs \be \eta(z)\eta(0)\sim -\fr{1}{2} \log z
\,,\qqu \td\eta(\zb)\td\eta(0)\sim -\fr{1}{2} \log \zb \,. \ee
Furthermore, any operator in the CFT with R-charges $(q,\tilde q)
=( \sqrt{k} p/2 , \sqrt{\tilde k}\tilde p/2)$ can be written
in the form \cite{Aharony:2001dp}
\begin{eqnarray}
  \label{eq:GeneralOp}
\mathcal {O}&=&e^{i(p
\eta+\tilde{p}\tilde{\eta})}P(\partial^n\eta,\bar{\partial}^{\tilde{n}}\tilde{\eta})\hat{\mathcal
{O}}\,,
\end{eqnarray}
where $P(\partial^n\eta,\bar{\partial}^{\tilde{n}}\tilde{\eta})$ is a
polynomial in any derivatives of $\eta$ and $\tilde{\eta}$, while
$\hat{\mathcal {O}}$ is an operator independent of
 $\eta$ and $\tilde{\eta}$.
  The exponential factor $e^{i(p
\eta+\tilde{p}\tilde{\eta})}$ has a non-trivial OPE with $J$ and
$\tilde{J}$, which induces the shift of conformal dimensions when we
deform the theory. In the following discussion, we  focus on the
scalar operators $Y_{p,\tilde{p}}\equiv e^{i(p
\eta+\tilde{p}\tilde{\eta})}$ which carry holomorphic and
anti-holomorphic dimensions
\be \label{DbarD}
\D= p^2/4 =q^2/k, \qquad\qquad \bar\D=  \tilde{p}^2/4 = \tilde q^2/\tilde k\,.
\ee
The relevant OPEs are
\begin{eqnarray}
  \label{eq:WaldId}
J(z)J(0)&\sim&\frac{k}{2z^2} \,,\nonumber\\
J(z)e^{i p \eta(0)}&\sim&\sqrt{ k}\frac{p}{2z} e^{i p
\eta(0)} = \fr{q}{z} e^{i p \eta(0)}\,.
\end{eqnarray}

Let us warm up by reviewing the calculation in \cite{Aharony:2001dp} for the lowest correction to the
conformal dimension from the SUSY breaking deformation
\begin{eqnarray}
  \label{eq:OPEfirst}
\delta_{h}\langle
Y_{p,\tilde{p}}(z,\bar{z})Y_{-p,-\tilde{p}}(0)\rangle&=&h\int
d^2 w\langle e^{ip\eta}(z)J(w)
e^{-ip\eta}(0)\rangle \langle
e^{i\tilde{p}\tilde{\eta}}(\bar{z})\tJ(\bar{w})
e^{-i\tilde{p}\tilde{\eta}}(0)\rangle\nonumber\\
&=&\frac{h q \td q}{z^{p^2/2}\bar{z}^{\tilde{p}^2/2}}\int
d^2w\lt|\frac{1}{w-z}-\frac{1}{w}\rt|^2\nonumber\\
&=&\frac{2\pi h q \td q}{z^{p^2/2}\bar{z}^{\tilde{p}^2/2}}\textrm{log}\frac{|z|^2}{|a|^2} \,,
\end{eqnarray}
where $a$ is the short-distance cutoff for the integral, which is absorbed by a
rescaling  of the operator $Y_{p,\tilde{p}}$. The correction to the
conformal dimension can be read from Eq.~(\ref{eq:OPEfirst}) as
\begin{eqnarray}
  \label{eq:ConDimFirst}
(-\pi h q \td q,-\pi h q \td q)\,.
\end{eqnarray}
The result agrees with the bulk calculation in Eq.~(\ref{eq:ChargeMassBdy3}). Here we see that the change of the conformal
dimension at the leading order in $h$ is proportional to the product of both
$U(1)$ R-charges.  Note that the shifts of holomorphic and anti-holomorphic scale dimensions are equal, so SUSY breaking does not change the spin of the operator.

As we will now show,  the conformal dimension is
modified at the second order in $h$ even if one of the $U(1)$ R-charges of the
operator vanishes. Without loss of generality, let us take
$\tilde{q}=0$.  We find
\begin{eqnarray}
  \label{eq:OPEsecond}
\delta_{\tilde{h}^2}\langle
Y_{p,0}(z,\bar{z})Y_{-p,0}(0)\rangle&=&\fr{h^2}{2}\int d^2 w_1d^2
w_2\langle e^{ip\eta}(z)J(w_1)J(w_2)e^{-ip\eta}(0)\rangle\langle \tJ(\bar{w}_1)\tJ(\bar{w}_2)\rangle\nonumber\\
&=&\frac{\tilde k h^2 q^2}{4z^{p^2/2}}\int
d^2w_1d^2w_2\(\frac{1}{w_1-z}-\frac{1}{w_1}\)\(\frac{1}{w_2-z}-\frac{1}{w_2}\)\frac{1}{(\bar{w}_1-\bar{w}_2)^2} \nn\\
&&\qqu\qqu+\langle Y_{p,0}(z,\bar{z})Y_{-p,0}(0)\rangle_0\ \delta_{h,2}\langle 1\rangle\nonumber\\
&=&-\frac{\pi \tilde k h^2 q^2}{4z^{p^2/2}}\int
d^2w_1d^2w_2\(\frac{1}{w_1-z}-\frac{1}{w_1}\)\frac{\delta^2(\vec w_2-\vec z)-\delta^2(\vec w_2)}{\bar{w}_1-\bar{w}_2} \nn\\
&&\qqu\qqu+\langle Y_{p,0}(z,\bar{z})Y_{-p,0}(0)\rangle_0\ \delta_{h,2}\langle 1\rangle\nonumber\\
&=&-\frac{\pi^2 \tilde k h^2 q^2}{2z^{p^2/2}}\textrm{log}\frac{|z|^2}{|a|^2}+\langle
Y_{p,0}(z,\bar{z})Y_{-p,0}(0)\rangle_0\
\delta_{h,2}\langle 1\rangle \,.
\end{eqnarray}
Here the last term, $\langle Y_{p,0}(z,\bar{z})Y_{-p,0}(0)\rangle_0\
\delta_{h^2}\langle 1\rangle$, indicates disconnected diagrams which
are canceled by vacuum corrections. Therefore we find that the
correction to the total conformal dimension, $\D_{\rm Tot}=\D +\bar\D,$ of $Y_{p,0}$ at the second order
is $\pi^2 \tilde k h^2 q^2/2$,  again with equal shifts in $\D$ and $\bar\D$.


\se{SUSY breaking to all orders in $h$} \la{se:allorders}

\begin{figure}[h]
\begin{center}
\includegraphics[width=0.25\textwidth]{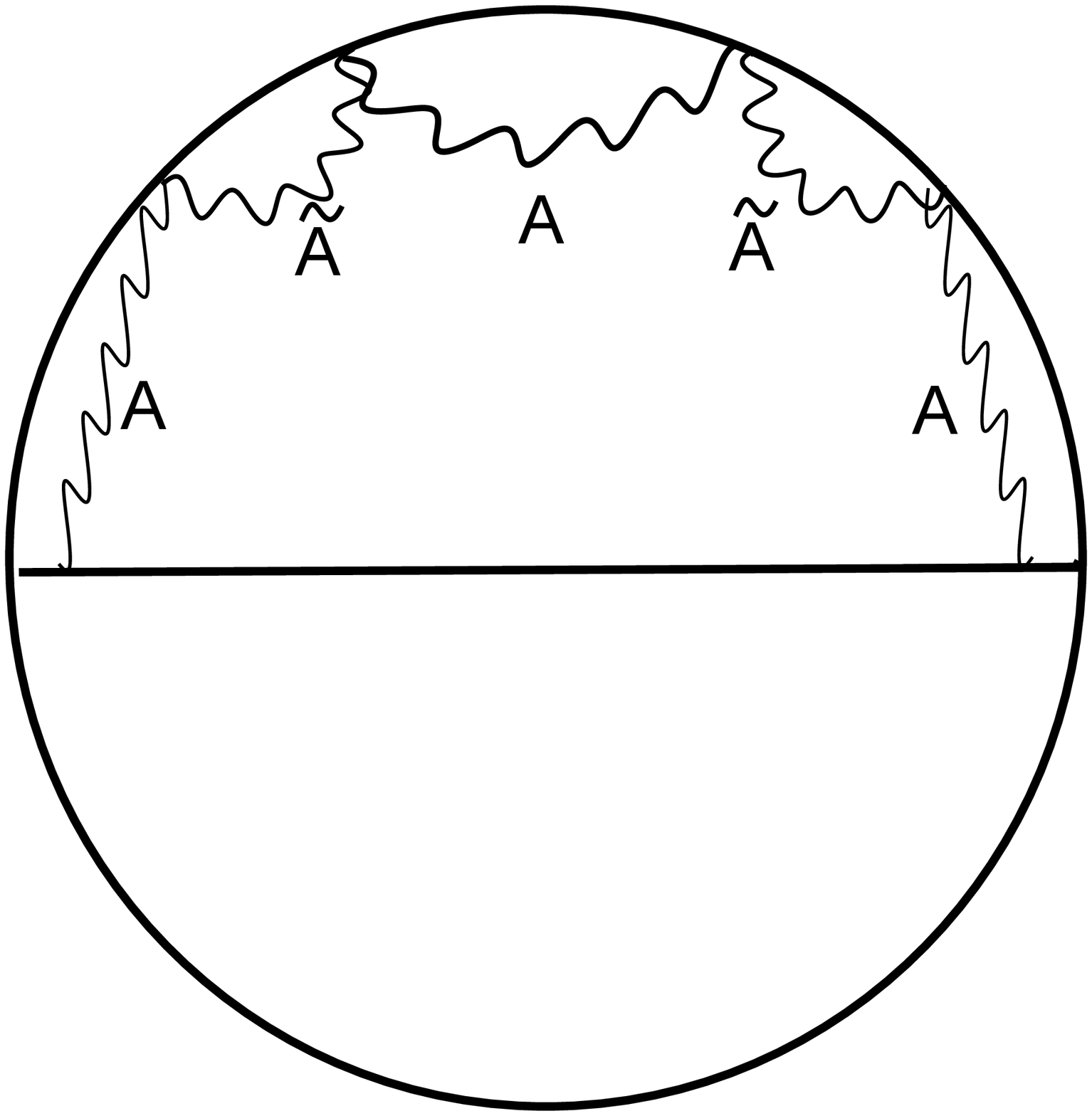}
\hspace*{0.35cm}
\includegraphics[width=0.25\textwidth]{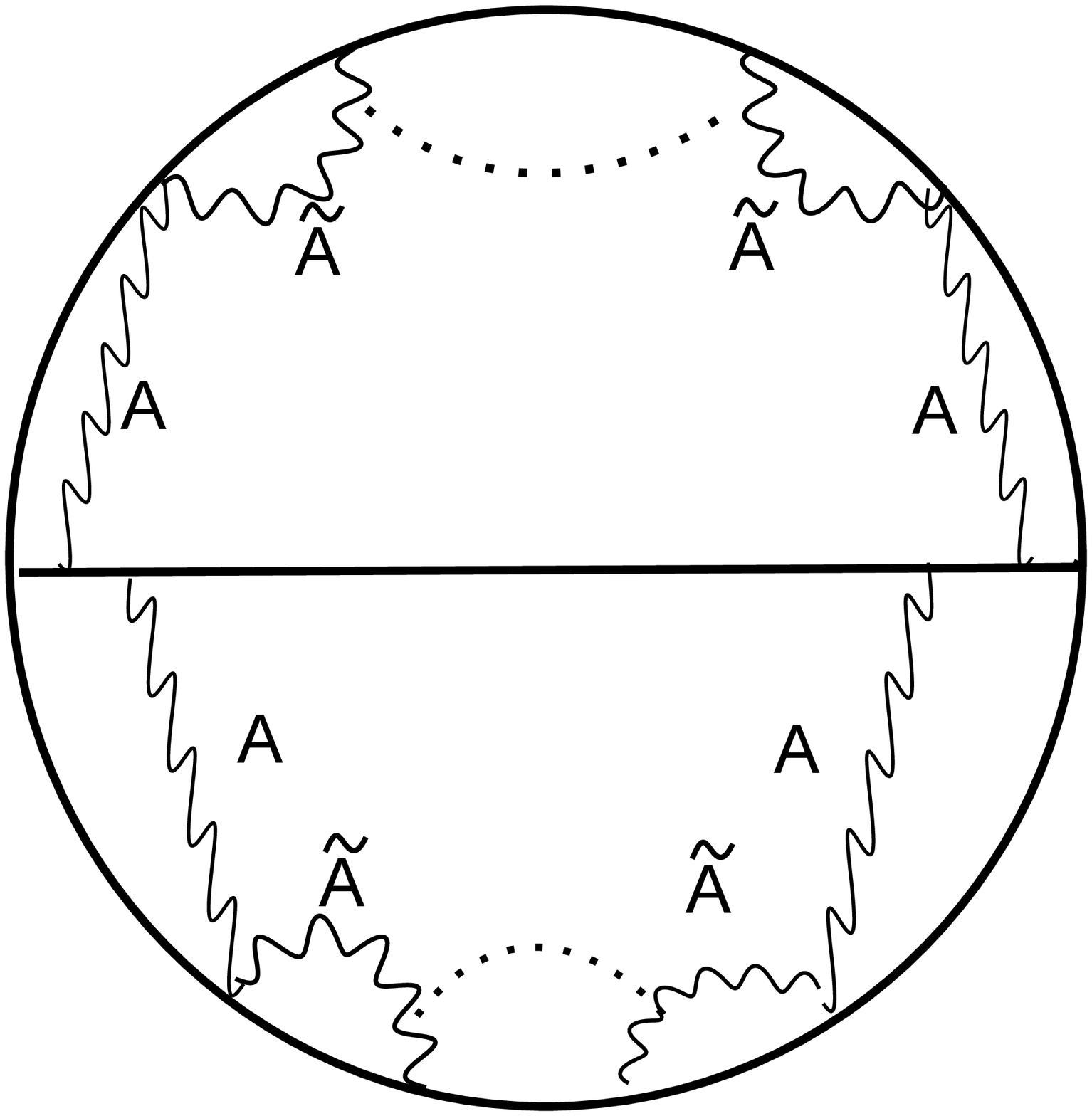}
\hspace*{0.35cm} \caption{Higher order SUSY breaking corrections to
the two-point correlation functions of R-charged particles. The
expansion is done for both $h$ and $q$. One first sums the higher
order $h$ expansion with a fixed order $q$, as shown on the left.
Then one can further sum the contributions on higher order $q$
expansion as shown on the right. } \label{fig:ChargedAll}
\end{center}
\end{figure}

The SUSY breaking shift of the scale dimension of an operator $\co_c$ due to interactions with the Chern-Simons fields was calculated to first order in $h$ for general $U(1)\times U(1)$ R-charges $(q,\tq)$ in Sec. 2.
 The result was confirmed by CFT methods in Sec. 4 and extended to second order.    In this section we return to the bulk theory and show that effects of the SUSY breaking can be summed to all orders in $h$.  We proceed in two stages:\\
i.)  The  sum of boundary insertions which "Wick contract" along the
boundary (See Fig.  \ref{fig:ChargedAll}) gives a "necklace"
structure which leads to a corrected  correlator of the form
 \be\label{necklace}
\<\co_c^\dagger\co_c\>   = \<\co_c^\dagger\co_c\>_0 \lt\{1+ \frac{2 \pi
hq\tq - \pi^2 h^2(\tilde k q^2+k \tq^2)/2}{1-\pi^2h^2 k\tilde
k/4}\log\frac{|y-z|^2}{a^2}\rt\}\,. \ee
ii.)   Further insertions of entire
necklaces can be summed to reveal that the single power of
$\log(|y-z|^2/a^2)$ is the beginning of an exponential series. The final form of the correlator is then the power law \be
\<\co_c^\dagger(y)\co_c(z)\> = C_0\frac{1}{|y-z|^{2\D}}\,, \ee
with
\be\label{exactdelta} \D =\D_{0} - \frac{ \pi hq\tq -
\pi^2h^2(\tilde k q^2+ k\tq^2)/2}{1-\pi^2h^2 k\tilde k/4}\,.
 \ee

We now show that the corrections to the holomorphic and anti-holomorphic parts of $\D_{\rm Tot}$  are positive for any operator of the form given in (\ref{eq:GeneralOp}) as required by unitarity.  We note that the undeformed $\D_0$ and $\bar\D_0$ are bounded below by $q^2/k$ and $\tq^2/\tilde k$, respectively. The bounds are saturated for the scalar operator $Y_{p,\tilde p}$.
Thus we can write
\be
\D\ge  q^2/k + \d \,,\qu
\bar\D \ge \tq^2/\tilde k + \d \,,\qu
\d = \frac{-\pi hq\tq + \pi^2h^2(\tilde k q^2+ k\tq^2)/4}{1-\pi^2h^2 k\tilde k/4} \,.
\ee
It is easy to see that the right sides of these inequalities are perfect squares, namely
\ba
q^2/k + \d &= \frac{(q - \pi hk \tq/2)^2 /k}{1-\pi^2h^2 k\tilde k/4} \,,\\
\tq^2/\tilde k + \d &= \frac{(\tq - \pi h\tilde k q/2)^2 /\tilde k}{1-\pi^2h^2 k\tilde k/4} \,.
\ea
Therefore after the deformation $\D$ and $\bar\D$ are manifestly non-negative. This satisfies the 2d unitarity bound (total dimension $\geq$ spin).

%
%
%
%

We now provide further details of the calculations that lead to the results above. In the diagrams to be evaluated, the bulk-to-boundary propagators of the gauge field $A(x),~\tilde A(x)$ are inserted are initially attached to the internal line of the bulk scalar field $\phi_c(x)$, and the Ward identity methods are applied with the result that the gauge fields are pinned at the boundary points $\vec y,~\vec z$.  We do not repeat these now-familiar arguments.

Feynman rules for the elements of the necklace diagrams in Fig.
\ref{fig:ChargedAll} may be obtained from (\ref{bulkdef}) and the
information in Appendix A.4. We use the boundary limits of
(\ref{holol}) and the limits recorded in (\ref{lim}).
\begin{eqnarray} \label{Feynrules}
{\rm internal~vertex} &:&  h\\
{\rm endpoint~attachment~of~A_i} &:& \frac{ q}{(y-w)}\\
{\rm endpoint ~attachment~of~\tilde A_i} &:& \frac{ \tilde q}{(\bar y -\bar w)}\\
{\rm internal~A_i ~line}&:& \frac{k}{2(w-w')^2}\\ 
{\rm internal~\tilde A_i ~line}&:& \frac{\tilde k}{2(\bar w-\bar
w')^2}
 \eea

Using the Feynman rules above, we can compare the bulk calculation of the
SUSY breaking correction for a field of charge  $(q,0)$ with the OPE calculation in (\ref{eq:OPEsecond}):
\begin{eqnarray}
\frac{h^2q^2\tilde k}{4|y-z|^{2\Delta_C}}\int d^2w_1d^2w_2\,
(\frac{1}{y-w_1}-\frac{1}{z-w_1})\frac{1}{(\bar w_1-\bar
w_2)^2}(\frac{1}{y-w_2}-\frac{1}{z-w_2})
\end{eqnarray}
Comparing with Eq. (\ref{eq:OPEsecond}), this agrees well with the
OPE calculation.

The summation of the beads of the necklace is facilitated by the
observation that the basic "unit" to be inserted in the transition
from order $h^{2(n-1)}$ to order $h^{2n}$ is the integral
 \bea \label{unit}
\frac{h^2 k \tilde k^2}{8}\int d^2ud^2u'
\frac{1}{(\bar w_{n-1}-\bar u)^2}\frac{1}{(u-u')^2}\frac{1}{(\bar
u-\bar w_{n})^2}
&=&\\
 \frac{h^2k \tilde k^2}{8} \int d^2ud^2u' \frac{\pa}{\pa u}\frac{1}{(\bar u-\bar w_{n-1})}\frac{\pa}{\pa \bar u}\frac{1}{(u-u')}\frac{1}{(\bar u-\bar w_{n})^2}&=&\\
 \frac{h^2 k \tilde k^2}{8} \int d^2ud^2u' \pi^2\d^{(2)}(u-w_{n-1})\d^{(2)}(u-u') \frac{1}{(\bar u-\bar w_{n})^2}&=&\\
\frac{\pi^2h^2 k \tilde k^2}{8}\frac{1}{(\bar w_{n-1} -\bar
w_{n})^2}\,. \eea The result is the insertion factor for an internal
$\tilde A_i$ line multiplied by the factor $\pi^2 h^2 k \tilde k/4$.
This leads to the geometric series  that is summed in
(\ref{necklace}).

For general charges $(q,\tq)$ one proceeds by similar methods. It is clear that the order $h^{2n}$ necklace diagrams are proportional to the factor $\tilde k q^2+k\tq^2$ and that order $h^{2n+1}$ contain the factor $q\tq$.  The necklace produces the same geometric series in both cases.  The result is given in (\ref{necklace}) above.

It is straightforward to understand the exponentiation that leads to
(\ref{exactdelta}). On the right in Fig. \ref{fig:ChargedAll} we
indicate the contribution of two complete necklace insertions.
Initially, the bulk-boundary propagators are coupled along the
charged line in the bulk.   Ward identity arguments apply to each
photon independently, so that each photon becomes pinned at the
boundary points $y,~z$ after all orders of attachment along the
charged line are added.  The result is the square of the single
necklace term in (\ref{necklace}) multiplied by the combinatoric
factor of 1/2.  Additional necklace insertions complete the
exponential series.

\se{Marginal deformation of $n$-point correlators }\label{sec:Deformation}

The diagrammatic calculations discussed in Secs.~\ref{se:ChargedTree}, \ref{se:ModuliMass}, and \ref{se:allorders} can be extended in a straightforward manner to $n$-point correlation functions.  The Ward identity ensures that  gauge fields that propagate from the boundary to each R-charged bulk line become pinned at the boundary points of that line.  This results in a simple all orders formula for the SUSY deformation of a general correlation function.

To discuss this formula it is useful to rewrite formula (\ref{exactdelta}) for the exact deformed scale dimension of an operator with R-charges $(q,\tq)$:
\bea
&&\D =\D_0 + A(h)q^2 +B(h)\tq^2 +C(h)q \tq\\
A(h)&=&\frac{\pi^2h^2\tilde k/2}{1-\pi^2h^2k\tilde k/4}\qquad B(h)=\frac{\pi^2h^2 k/2}{1-\pi^2h^2k\tilde k/4}\qquad C(h) = \frac{2\pi h}{1-\pi^2h^2k\tilde k/4}
\eea
We now consider an $n$-point correlator of operators $O_{(q_i,\tilde q_i)}$.  Ward identity arguments imply that the exact relation between the deformed and undeformed correlators is (with $\vy_{ij} =\vy_i-\vy_j$)
\be\label{deform1}
\<O_{(q_1,\tilde q_1)}(\vy_1)\ldots O_{(q_n,\tilde q_n)}(\vy_n)\> \sim \<O_{(q_1,\tilde q_1)}(\vy_1)\ldots O_{(q_n,\tilde q_n)}(\vy_n)\>_{{}_0}\prod_{i\ne j} |\vy_{ij}|^{A(h)q_iq_j+B(h)\tq_i\tq_j-C(h)q_i\tq_j}
\ee
where $\sim$ indicates that the formula holds up to a dimensionful constant.
 Since the formula emerges from an exactly marginal deformation of a CFT, we know in advance that the deformed correlator transforms properly under conformal transformations.
Nevertheless, it is curious and satisfying to check that the right side transforms with appropriate weights under inversion,  $\vy_i = \vy'_i/(\vy'_{ij})^2$.  One finds that it does transform with deformed weights for all operators, provided that R-charge is conserved, i.e. $\sum_i q_i =\sum_i \tq_i =0.$ Thus charge conservation is linked to conformal invariance.

There are  further, equally simple formulas for correlators involving R-charged operators together with currents or the stress tensor. For example, with one additional R-current added, we have the formula
\be\label{deform2}
\<J_i(\vx) O_{(q_1,\tilde q_1)}(\vy_1)\ldots O_{(q_n,\tilde q_n)}(\vy_n)\> \sim \<J_i(\vx) O_{(q_1,\tilde q_1)}(\vy_1)\ldots O_{(q_n,\tilde q_n)}(\vy_n)\>_{{}_0}\prod_{i\ne j} |\vy_{ij}|^{(A(h)q_iq_j+B(h)\tq_i\tq_j-C(h)q_i\tq_j)}\,.
\ee
Note that the Ward identity is satisfied by the deformed correlator simply because the deformation does not change the dependence on $\vx$.

\se{Correction to bulk coupling constants}

In this section, we show that the coupling constants in the bulk
generally shift once we turn on the deformation \er{action}.  For
concreteness let us focus on the cubic coupling $\l_3 \p_c^\dag \p_c
\p_m$, where $\p_c$ denotes a scalar field with R-charges $(q,\td
q)$ and $\p_m$ denotes a neutral scalar field such as a modulus.  We
will use the change of the 3-point function $\<\co_c^\dag \co_c
\co_m\>$ to determine the correction to the cubic coupling.

\sse{Correction to the 3-point function}

In this subsection we calculate the correction to the 3-point
function $\<\co_c^\dag \co_c \co_m\>$.  By conformal invariance it
must have the following structure in the undeformed theory: \be
\<\co_c^\dag(\vec y) \co_c(\vec z) \co_m(\vec w)\> = \fr{c_3}{|\vec
y-\vec z|^{2\D_c-\D_m} |\vec y-\vec w|^{\D_m} |\vec z-\vec
w|^{\D_m}} \,. \ee When we turn on $h$, both $c_3$ and $\D_c$
change, but we also need to correctly normalize the 2-point function
of $\co_c$.  According to \er{eq:ChargeMassBdy3}, the 2-point
function to the first order in $h$ is \be\la{ooh} \<\co_c^\dag(\vec
y) \co_c(\vec z)\>_h = \fr{2(\D_c-1)^2}{\pi}\fr{a^{-2\b h}}{|\vec
y-\vec z|^{2(\D_c-\b h)}} \,, \ee where we have defined $\b=2\pi q
\td q$, and $a$ is the short distance cutoff.  Therefore, the
correction to $\D_c$ is $-\b h$, and we define \be \td\co_c = \co_c
a^{\b h} \(1 - \fr{\b h}{\D_c-1}\) \,, \ee so that the 2-point
function for $\td\co_c$ is properly normalized: \be
\<\td\co_c^\dag(\vec y) \td\co_c(\vec z)\>_h = \fr{2(\D_c-\b
h-1)^2}{\pi}\fr{1}{|\vec y - \vec z|^{2(\D_c-\b h)}} \,. \ee We
recall that proper normalization of  the 2-point function corresponds to
canonical normalization of the dual scalar field in the bulk.

By essentially the same calculation (performed either in the bulk or
using OPE techniques in the CFT) that led to \er{ooh}, we find that
the 3-point function $\<\co_c^\dag \co_c \co_m\>$ to first order
in $h$ is \be \<\co_c^\dag(\vec y) \co_c(\vec z) \co_m(\vec w)\>_h =
\fr{c_3 a^{-2\b h}}{|\vec y-\vec z|^{2(\D_c-\b h)-\D_m} |\vec y-\vec
w|^{\D_m} |\vec z-\vec w|^{\D_m}} \,, \ee where $c_3$ and $\D_c$ are
defined in the undeformed theory, and we have indicated the
correction of order $h$ explicitly. Therefore, written in the
properly normalized $\td\co_h$, we have \be\la{ototom}
\<\td\co_c^\dag(x_1) \td\co_c(x_2) \co_m(x_3)\>_h = \fr{c_3 \(1 -
\fr{2\b h}{\D_c-1}\)}{|\vec y-\vec z|^{2(\D_c-\b h)-\D_m} |\vec
y-\vec w|^{\D_m} |\vec z-\vec w|^{\D_m}} \,. \ee From this we can
extract the first-order correction to the coefficient of the
properly normalized 3-point function: \be\la{dc3} \d c_3 = - \fr{2\b
h}{\D_c-1} c_3 \,. \ee

In an AdS/CFT calculation, the 3-point function is determined in
terms of the bulk cubic coupling $\l_3$ and the integral of a
product of three bulk-to-boundary propagators: \be\la{ooo}
\<\co_c^\dag(\vec y) \co_c(\vec z) \co_m(\vec w)\> = -\l_3 \int
\fr{d^3x}{x_0^3} K_{\D_c}(x,\vec y) K_{\D_c}(x,\vec z)
K_{\D_m}(x,\vec w) \,. \ee As we turn on the $h$ deformation, both
the cubic coupling $\l_3$ and the bulk-to-boundary propagator
$K_{\D_c}$ change.  In order to determine the correction to $\l_3$,
we next calculate how $K_{\D_c}$ changes.

\sse{Correction to the bulk-to-boundary propagator}

We now calculate the correction to the bulk-to-boundary propagator
$K_{\D_c}$ by first computing the corrected bulk propagator to the
first order in $h$.  After using the same argument that involves
integration by parts and led to Sec.~\re{se:ChargedTree}, we find
the first-order correction to the bulk propagator is \be \d_h
G_{\D_c}(x,x') = h q \td q G_{\D_c}(x,x') \int d^2w \[\L_z(x,\vec
w)-\L_z(x',\vec w)\] \[\td\L_{\zb}(x,\vec w)-\td\L_{\wb}(x',\vec
w)\] \,. \ee Performing the integrals, we find \be\la{dgdc} \d_h
G_{\D_c}(x,x') = \b h G_{\D_c}(x,x') \lt\{\fr{u+1}{\sqrt{u(u+2)}}
\log\[u+1 +\sqrt{u(u+2)}\]-1\rt\} \,, \ee where $u$ is the
bi-invariant variable defined in \er{udef}.  Let us also recall the
$D=3$ bulk propagator in the undeformed theory \be\la{gdc}
G_{\D_c}(x,x') = \fr{2^{\D_c-2}}{\pi} \(\sqrt{u}
+\sqrt{u+2}\)^{-2\D_c} \[1+ \fr{u+1}{\sqrt{u(u+2)}}\] \,. \ee

We find the corrected bulk-to-boundary propagator $K_{\D_c, h}$ by
taking a limit of the corrected bulk propagator $G_{\D_c, h} =
G_{\D_c} + \d_h G_{\D_c}$: \be\la{kdch} K_{\D_c, h}(x,\vec x') \sim
\lim_{x_0'\to 0} x_0'^{-(\D_c-\b h)} G_{\D_c, h} (x,x') \,, \ee
where we have used the fact that the corrected dimension is $\D_c-\b
h$.  The normalization for $K_{\D_c, h}$ is not specified above, but
is easily determined by the usual boundary condition \be
\lim_{x_0\to 0} x_0^{\Delta_c-\b h-2} K_{\Delta_c,h}(x,\vec x') =
\delta^2(\vec x,\vec x') \,. \ee

Plugging \er{dgdc} and \er{gdc} into \er{kdch}, we find \be K_{\D_c,
h}(x,\vec x') = K_{\D_c-\b h}(x,\vec x') = \fr{\D_c-\b h-1}{\pi}
\(\frac{x_0}{x_0^2 + (\vec x-\vec x')^2 }\)^{\D_c-\b h} \,. \ee In
other words, the correction to the bulk-to-boundary propagator is
exactly accounted for by replacing $\D_c$ with the corrected
dimension $\D_c -\b h$ in the bulk-to-boundary propagator of the
undeformed theory.

\sse{Correction to the cubic coupling}

Now that we understand the correction to both the 3-point function
and the bulk-to-boundary propagator, we can calculate the correction
to the cubic coupling $\l_3$.  Assuming that the 3-point function
$\<\co_c^\dag \co_c \co_m\>$ is completely determined from \er{ooo},
we can evaluate the integrals there and find \cite{Freedman:1998tz}
\be c_3 = -\l_3 \fr{\G\(\D_c -\fr{\D_m}{2}\) \G\(\fr{\D_m}{2}\)^2
\G\(\D_c +\fr{\D_m}{2} -1\)}{2\pi^2 \G\(\D_c-1\)^2 \G\(\D_m-1\)} \,.
\ee As we turn on the $h$ deformation, $c_3$, $\l_3$, and $\D_c$ all
receive corrections but continue to satisfy the above equation.
Using $\d_h c_3 = -\fr{2\b h}{\D_c-1} c_3$ and $\d_h \D_c = -\b h$,
we find \ba
\fr{\d_h \l_3}{\l_3} &= -\fr{2\b h}{\D_c-1} + \b h \fr{\pa}{\pa\D_c} \log \fr{\G\(\D_c -\fr{\D_m}{2}\) \G\(\D_c +\fr{\D_m}{2} -1\)}{\G\(\D_c-1\)^2} \\
&= \b h \fr{\pa}{\pa\D_c} \log \fr{\G\(\D_c -\fr{\D_m}{2}\) \G\(\D_c
+\fr{\D_m}{2} -1\)}{\G\(\D_c\)^2} \,. \ea Our calculation applies to
any $\co_m$ that is neutral under the R-symmetry group.  However, we
are perhaps most interested in the case where $\co_m$ is a modulus.
This means $\D_m=2$, and the above formula simplifies to \be \lt.
\fr{\d_h\l_3}{\l_3} \rt|_{\D_m=2} = -\fr{\b h}{\D_c-1} = -\fr{2\pi h
q \td q}{\D_c-1} \,. \ee where we have used $\b=2\pi q \td q$. This
is an interesting result that depends on the R-charges $(q, \td q)$
and the dimension $\D_c$ of the scalar field $\p_c$.  In particular,
this means that the supersymmetric relations between coupling
constants in the undeformed theory are generally broken by the
deformation.

In general, the 3-point function $\<\co_c^\dag \co_c \co_m\>$ might
not be completely determined by the single cubic coupling $\l_3
\p_c^\dag \p_c \p_m$ via the AdS/CFT calculation \er{ooo}; for
example, higher-derivative bulk couplings such as $\l_3' \pa^\m
\p_c^\dag \pa_\m \p_c \p_m$ also contribute to the same 3-point
function if they exist in the theory.  Therefore, the change of the
3-point function coefficient \er{dc3} may be attributed to
corrections to both $\l_3$ and its higher-derivative cousins such as
$\l_3'$.  We expect that a careful analysis of 4-point functions may
unambiguously determine the corrections to all these coupling
constants separately, and leave this to future work.  For our
current purposes, it is sufficient to show that the coupling
constants in the bulk generally receive corrections from our
deformation, and the supersymmetric relations between them in the
undeformed theory are generally broken.

\section{The deformation is exactly marginal}\la{se:mg}

Our main purpose in this section is to exhibit the marginal property
in terms of bulk calculations,  but we begin with a brief summary of the CFT
result of \cite{Chaudhuri:1988qb}.  The authors consider a set of
holomorphic currents  $J^a(z)$  of conformal dimension $(1,0)$
which enjoy the usual OPE of a current algebra, namely \be
\label{usualope} J^a(z)J^b(w) \sim \frac{k^{ab}}{(z-w)^2}
+i \frac{f^{abc}}{z-w} J^c(w)\,, \ee together with a
similar set  of anti-holomorphic $\tilde J^a(\bar z)$. They then
prove that an operator of the bilinear form
\be
O(z,\bar z) = c_{ab} J^a(z)\tilde J^b(\bar z)
\ee
is exactly marginal if and only if it can be rewritten in the form
\be \label{margop}
O(z,\bar z) = c'_{ab} V^a(z)\tilde V^b(\bar z)
\ee
where the $V^a$ (or $\tilde V^b$) operators are linear combinations of the $J^a$ (or $\td J^b$) currents and there is no simple pole in the OPE among the $V^a$ (and $\td V^b$) operators. Since our deformation is a product of two abelian
currents, i.e. $O(z,\bar z)  = J(z)\tilde J(\bar z)$, it satisfies
this condition quite trivially.

Let us now turn to the bulk theory and exhibit the exact marginality of our deformation $O(z,\zb)=J(z)\td J(\zb)$ there.  We evaluate the Witten diagrams that contribute to the 2-point function $\<OO\>$
using Wick contractions. Let $y,~z$ be the boundary points. The
basic Wick contractions are obtained from an  argument similar to the one leading to \er{2ptj}: \be \label{wick} [A_y A_z]  =
\frac{k}{2(y-z)^2} \,,\qquad\qquad [\tA_{\bar y}\tA_{\bar z}]=
\frac{k}{2(\bar y-\bar z)^2}\,. \ee We use $[\ldots]$ to
indicate Wick contractions. Then the undeformed correlator is given by the Wick contraction: \be
\langle O(y,\bar y)O(z, \bar z)\rangle_0 = [A_y\tA_{\bar y} A_z \tA_{\bar z}] =
\frac{k^2}{4|y-z|^4}\,. \ee
We now test whether corrections due to
the boundary deformation contain logarithmic terms that indicate a shift of the conformal dimension.  At the first order in $h$
we encounter the contractions in \be \langle O(y,\bar y)O(z, \bar z)\rangle_1 = h
\int d^2w [A_y\tA_{\bar y} A_w\tA_{\bar w} A_z \tA_{\bar z}]\,. \ee But
the net contraction among three (or any odd number of) $A$'s
vanishes.  Hence there is no correction at order $h$. At the next order we need
to consider \be \langle O(y,\bar y)O(z, \bar z)\rangle_2 = \fr{h^2}{2} \int d^2w_1\,d^2w_2[
A_y\tA_{\bar y} A_{w_1}\tA_{\bar w_1} A_{w_2}\tA_{\bar w_2} A_z
\tA_{\bar z} ]\,. \ee There are several inequivalent products of
four contractions each, and each product corresponds to a distinct
Witten diagram as shown in Fig.~\ref{fig:App}.

\begin{figure}[h]
\begin{center}
\centering
\includegraphics[width=0.25\textwidth]{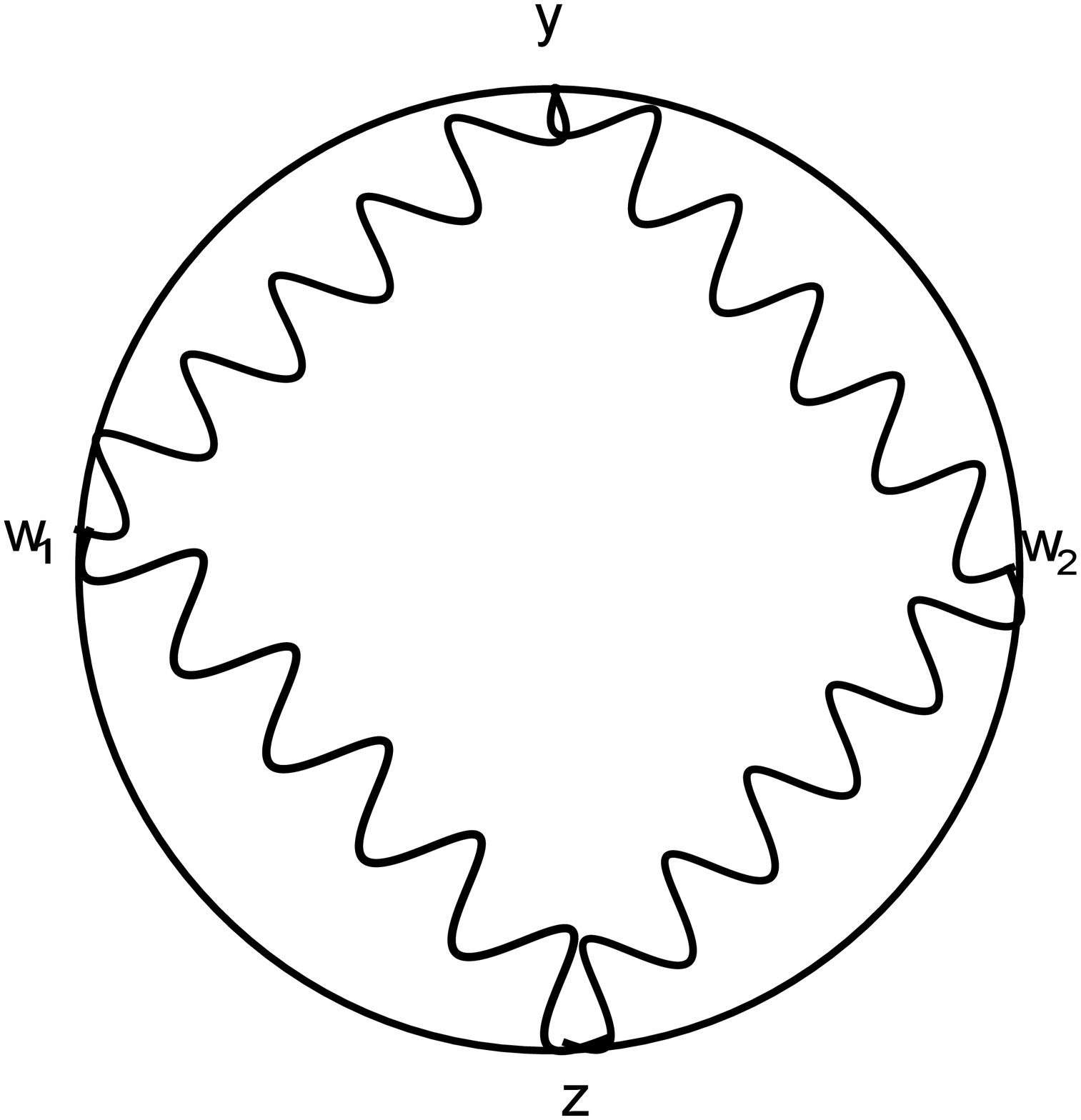}$(a)$
\hspace*{0.35cm}
\includegraphics[width=0.25\textwidth]{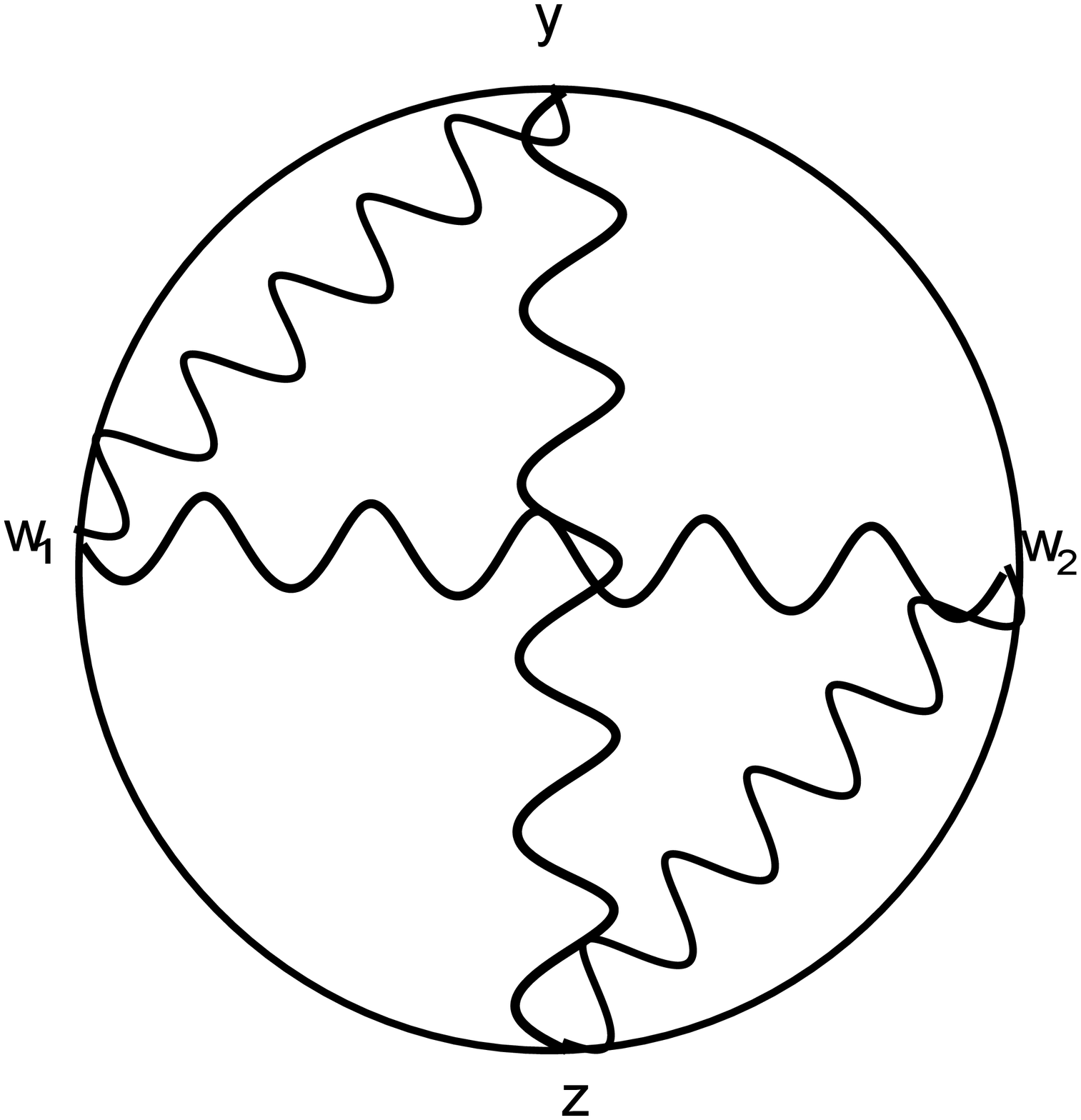}$(b)$
\hspace*{0.35cm}
\includegraphics[width=0.25\textwidth]{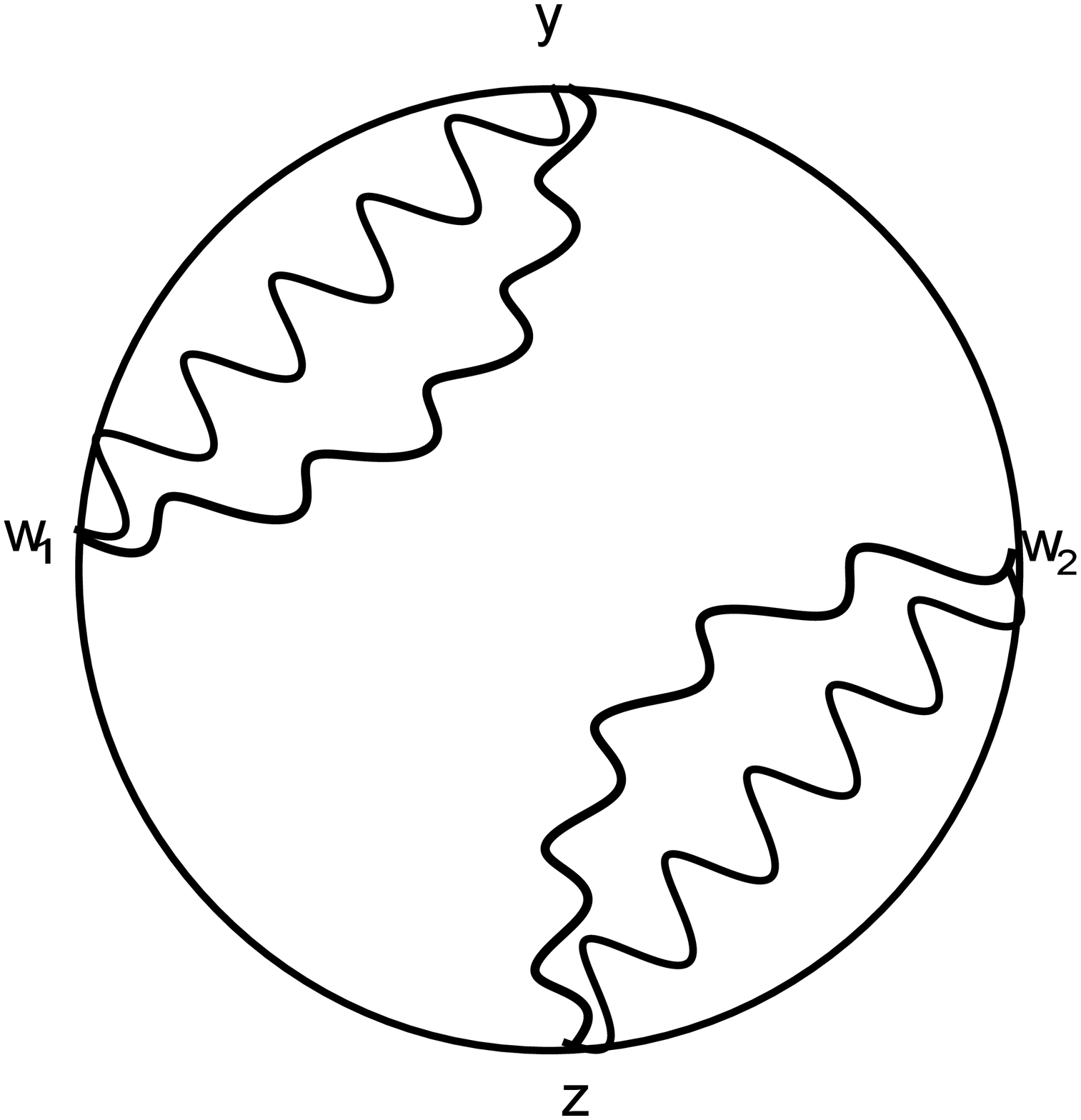}$(c)$
\hspace*{0.35cm}
\includegraphics[width=0.225\textwidth]{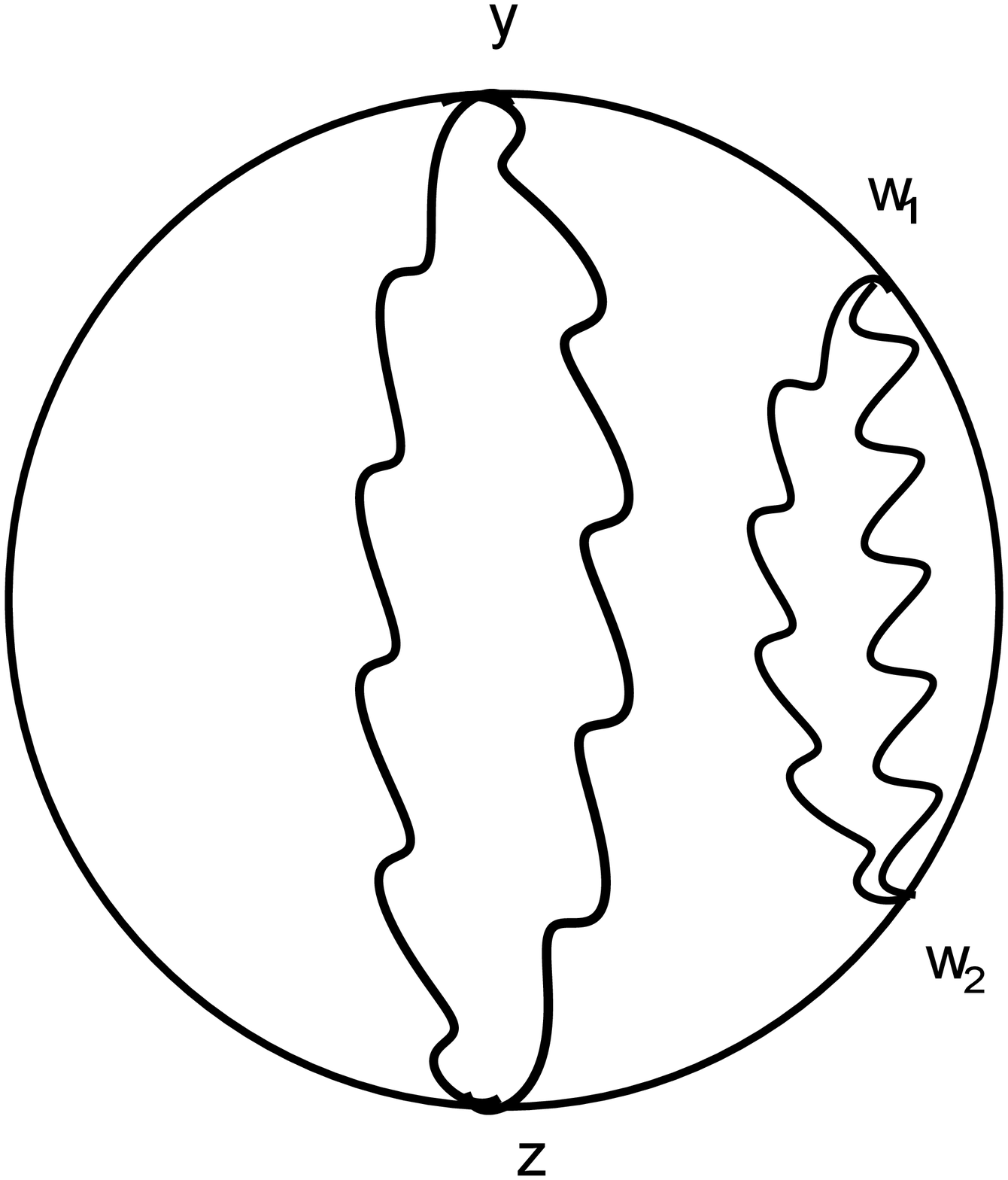}$(d)$
\hspace*{0.35cm}
\includegraphics[width=0.25\textwidth]{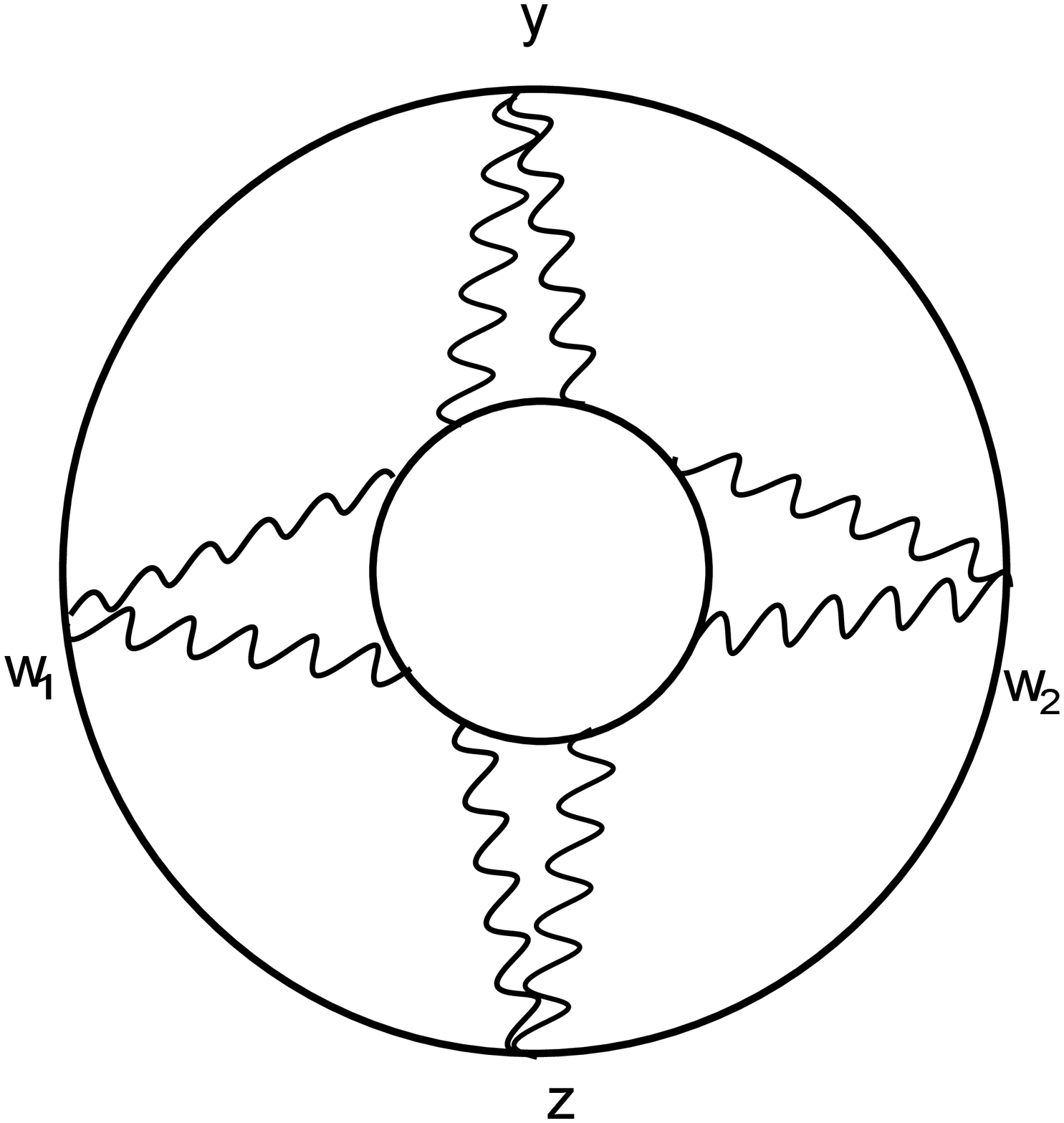}$(e)$
\caption{Diagrams relevant to the exactly marginal property of the deformation.}
 \label{fig:App}
\end{center}
\end{figure}

We will not present   details,  but simply note that the Wick contractions are purely holomorphic or anti-holomorphic.  Thus standard CFT techniques can be used to evaluate the $d^2w_1d^2w_2$ integrals. Below is the result for each of the diagrams.

Fig.~\ref{fig:App}a:  This gives a contact term of no interest since we are concerned with the correlator for $x \ne y$.

Fig.~\ref{fig:App}b:  This gives  finite term proportional to $1/|y-z|^4$ which corrects the normalization of the correlator but not the conformal dimension of $O$.

Fig.~\ref{fig:App}c:  This gives the product of two divergent
one-point functions.  The divergence can be cancelled by
counterterms.

Fig.~\ref{fig:App}d:  This disconnected diagram gives a divergent
result, but as usual it cancels with vacuum corrections and does not contribute to the correction of the 2-point function.

Fig.~\ref{fig:App}e: This is one of several diagrams that contain a charged particle loop. These diagrams vanish by the argument applied to moduli fields in Sec.~\ref{se:ModuliMass}.

This argument shows that the conformal dimension (and hence the marginality) of our deformation operator $O(z,\zb)=J(z)\td J(\zb)$ is not modified when we turn on the $h$ deformation, through cubic order in $h$.  The argument can be extended to all orders in $h$ as in Sec.  \ref{se:allorders}.

\section{Global SUSY in the undeformed theory}

Supersymmetry of the undeformed bulk theory is an important element
of our work,  but it has not been explored directly in any of the
calculations described above.  Suppose for example that  $\phi$ and
$\chi$ are the scalar and spinor components of a chiral multiplet in
$AdS_3$. In this section we show that these quantities and their
masses are properly related by $AdS$ supersymmetry.  The argument
will clarify the nature of the mass term required in the
supergravity theory that underlies our work.

It is reasonably well known that the mass parameters for scalars and
spinors in a chiral multiplet are not equal in a supersymmetric
field theory in $AdS$.  The conventional mass term in Euclidean
$AdS_3$ is \be \label{Lmcon} \cl_{\m} = \frac{1}{L^2}\[\(-\frac{3}{4}
+\m^2\)\phi^\dag\phi + \m\(\phi^2+\phi^{\dag2}\)  -\frac{\m
L}{2}\(\chi^2 +\bar\chi^{2}\)\]\,. \ee The $\mu$ parameter here is
supersymmetric, but measured in units of $1/L$. It can be thought of
as descending from the superpotential $W= \m \phi^2/(2L)$ in $D=4$.
This mass term obviously does not conserve the $R$-charge, so it is
inadmissible in our present theory in which the $R$-charge is gauged by
Chern-Simons fields.

Fortunately there is an alternative mass term, called the ``real
mass,'' which is special to three-dimensional SUSY.  The key feature
is that the mass parameters are related to the $R$-charge of the
multiplet.  For simplicity we assume that the scalar $\phi$ carries
R-charges $(q,0)$. We obtain the real mass term for Euclidean $AdS_3$
from the $S^3$ version of Jafferis \cite{Jafferis:2010un} by the
replacement $a\to i L$, where $a$ is the radius of the sphere: \be
\label{Lmjaff} \cl_q = \frac{1}{L^2}\[\(-\frac{3}{4} +\(q-\fr12\)\(q-\fr32\)\)
\phi^\dag\phi- i \(q-\fr12\)L\bar\chi\chi\]\,. \ee This conserves
the $R$-charge! It is admissible in our framework and indeed required by
SUSY as we now show.

We see that mass parameters $m_B^2$ and $m_F$ of the scalar $\phi$ and fermion $\chi$ in a chiral supermultiplet are related to the
$R$-charge $q$ by
\be \label{mq} (m_B L)^2 = -\frac34 +
\(q-\fr12\)\(q-\fr32\) \,,\qquad\quad m_F L = q-\fr12\,.
 \ee
It follows from the $AdS_3$ supersymmetry algebra that
$\phi$ and $\chi$ have conformal dimensions $\D_B$ and $\D_F$ related by
$\D_F = \D_B + 1/2.$   Finally we write the AdS/CFT formulas that
relate conformal dimensions to Lagrangian mass parameters by
\be\label{deltascale} \D_B = 1 + \sqrt{1+(m_BL)^2} \,,\qquad\quad \D_F =
1 +|m_FL|\,. \ee

We want to show that the effect of a small
supersymmetric variation of these quantities is consistent with the
mass relations of \er{mq}.
 ~Therefore we compute the variations
\be\label{vars1}
\d(m_B L)^2 = 2(q-1)\d q\,,\qquad\quad \d(m_FL) = \d q \,.
\ee
To maintain the  supersymmetry relation $\D_F = \D_B + 1/2$, we require $\d\D_F =\d\D_B$.  Hence we test this:
\bea
\d\D_F &=& \d q \,,\\
 \d\D_B&=&\frac{\d (m_BL)^2}{2 \sqrt{1+(m_BL)^2}} = \frac{(q-1)\d q}{\sqrt{(q-1)^2}}= \d q\,.
 \eea
We pass this test and thus verify that the mass parameters of \er{Lmjaff} are consistent with SUSY.
\section{Hierarchy and little hierarchy problem in 3d}

Many physicists favor supersymmetry as the solution to the hierarchy problem in particle physics.
The
introduction of superpartners of all standard model (SM) particles
cancels  quadratically divergent quantum corrections to the Higgs mass.
When SUSY is spontaneously broken,  mass differences between SM particles and their
superpartners are generated. Consequently, the mass of the Higgs boson will be corrected due to the mismatch of particle spectra and the running of
coupling constants below the SUSY breaking scale. The lack of evidence
for superpartners in the LHC data below its present limit of order TeV implies that a sizable fine tuning is needed in the MSSM to account for the low electroweak scale. This is the little
hierarchy problem.

Our aim in this paper is to find a SUSY breaking mechanism which can
induce sizable mass splitting in supermultiplets while still
protecting light scalar masses from quantum corrections. Our toy
model is a field theory living in the $AdS_3$ spacetime. Thus we
would  like to show that there is a hierarchy problem in a generic
$D=3$  theory, and a little hierarchy problem after SUSY breaking in
a SUSY theory. Since this question concerns UV physics, we work in
flat $D=3$ spacetime for simplicity.

Let us start with a simple four dimensional SUSY model, and write the
superpotential as
\begin{eqnarray}
  \label{eq:W4D}
W_{4D}=\frac{y}{3}\Phi^3+\frac{y'}{3M}\Phi^3\Phi'
\end{eqnarray}
where the $R$-charge for $\Phi$ is $\frac{2}{3}$ and $R$-charge for
$\Phi'$ is 0. $\frac{y'}{3M}\Phi^3\Phi'$ is an irrelevant operator,
and $M$ is its suppression scale. The couplings $y,~y'$ are dimensionless. The Lagrangian induced by this
superpotential is
\begin{eqnarray}
  \label{eq:L4D}
L_{4D}&\supset & \(y^*\phi^{\dag 2}+\frac{y'^{*}}{M}\phi^{\dag
2}\phi'^\dag\)\(y\phi^2+\frac{y'}{M}\phi^2\phi'\)+\frac{y'^*
y'}{9M^2}(\phi^\dag\phi)^3\nonumber\\
&&\qqu +2y\phi\psi^2+\frac{2y'}{M}\phi\phi'\psi^2+\frac{y'}{M}\phi^2\psi\psi'+h.c.
\end{eqnarray}
We compactify this model on a circle of circumference $R$ to obtain a 3-dimensional supersymmetric theory. Only  zero modes contribute to the low
energy effective theory. Compactification
introduces an overall factor of $R$ in the $D=3$ Lagrangian, and the Lagrangian can be
properly normalized by scaling both scalar and
fermion fields by a factor of $\sqrt{R}$.  We take $R= 1/M$ for simplicity and write
the  $D=3$ interaction Lagrangian as
\begin{eqnarray}
  \label{eq:L3D}
L_{3D}&\supset& (\sqrt{M} y^*\phi^{\dag 2}+y'^{*}\phi^{\dag 2}\phi'^\dag)(\sqrt{M}y\phi^2+y'\phi^2\phi')+(y'^*
y'/9)(\phi^\dag\phi)^3 \nonumber\\
&&\qqu+ 2\sqrt{M} y\phi\psi^2+2y'\phi\phi'\psi^2+y'\phi^2\psi\psi'+h.c.
\end{eqnarray}
Note that the dimensions of $\phi$ and $\psi$ are $\frac{1}{2}$ and
$1$ respectively, as appropriate for $D=3$.  Furthermore, $L_{3D}$ contains only  marginal and relevant operators.

There are several ways to generate quantum corrections to the scalar
mass. For example,  at the 2-loop order the self-contractions of the
marginal operator $\phi^6$  produce quadratic divergences, and the
contractions between a pair of quartic $\phi^4$  give log
divergences. When SUSY is not broken, the corresponding fermionic
diagrams precisely cancel these divergences. The cancellations
require both the matches of particle spectra and coupling constants.
If SUSY is spontaneously broken, then below the SUSY breaking scale,
the mismatch of boson/fermion spectra as well as the running of
coupling constant induce uncancelled contributions to the scalar
mass. This is precisely the 3d analogue of the little hierarchy
problem in the standard model.

\section{Discussion}
In this paper, we find a novel SUSY breaking mechanism which may
shed some light on
 the solution of the (little) hierarchy problem in the MSSM. We
start from a supergravity theory with Chern-Simons gauge fields in
$AdS_3$. These fields gauge a $U(1) \times \tilde U(1)$ R-symmetry.
Then we introduce an explicit SUSY breaking boundary term quadratic
in these gauge fields. The SUSY breaking effects propagate to the
bulk through gauge couplings. As a result, the SUSY relation between
masses of bosons and fermions in each supermultiplet is violated. The
coupling constants of interaction vertices are also modified.
However, moduli fields, which are neutral under Chern-Simons gauge
transformations, maintain their flat potential to all orders in
perturbation theory.

This is a surprising result because one generically expects SUSY
breaking effects to migrate to  gauge neutral fields through quantum
loop corrections. We provide a comprehensive analysis to show that
such SUSY breaking effects are blocked.  It relies on the fact that
the bulk-to-boundary propagator of the Chern-Simons gauge fields is
a total derivative with respect to the bulk coordinates. Using
integration by parts and the Ward identity, one can easily prove
that SUSY breaking effects precisely cancel within charged loop
diagrams when calculating the quantum corrections to the potentials
of the moduli fields. From the effective field theory point of view,
there are two kinds of changes in the quantum loop corrections.
Internal propagators of charged particles are modified by the SUSY
breaking deformation and coupling constants of interaction vertices
are also changed. The effects from these two kinds of changes
precisely cancel and leave the potential of the moduli fields flat.

In the MSSM, a conventional way to estimate the amount of fine tuning is
first to calculate the stop particle's loop corrections to the soft SUSY breaking
mass $m_{H_u}^2$, i.e.
\begin{eqnarray}
  \label{Conventional}
\delta m_{H_u}^2&=&
-\frac{3y_t^2}{4\pi^2}m_{\tilde{t}}^2 \log\(\frac{\Lambda_{UV}}{m_{\tilde{t}}}\) \,.
\end{eqnarray}
Then one compares the soft mass corrections with the electroweak
scale to obtain the fine tuning \cite{Brust:2011tb,Craig:2013cxa}.
However, our toy model shows that this conventional estimate of
fine tuning may not provide the correct intuition when the complete UV
physics is unknown. Specifically, a mismatch of the masses within a supermultiplet does not always imply a mass correction to other fields.

The primary ``observables'' in AdS/CFT are the correlation functions of the boundary CFT.  In this viewpoint the $AdS_3$ analogue of the hierarchy problem is solved in the model that we present here.  It is, however, worth exploring the bulk physics in more detail.  Is bulk locality preserved by the SUSY breaking boundary term?  Is there a well-defined flat spacetime limit in which SUSY breaking effects persist?  Further investigation is needed to answer these open questions.

\section*{Acknowledgement}
We thank Michael Crossley, Savas Dimopoulos, Jean-Francois Fortin,
Shamit Kachru, Stephen Shenker, and  Eva Silverstein for helpful
discussions.  XD is supported by the National Science Foundation
under grant PHY-0756174. The research of DZF is supported in part by
NSF grant PHY-0967299. YZ is supported by ERC grant BSMOXFORD no.
228169.

\appendix

\section{Appendix:  Chern-Simons propagators in AdS/CFT}\la{ap:prop}

The ``pure gauge'' structure of the bulk-to-boundary propagator, i.e.
$K_{\m i'}(x,\vec x') = \pa_\m \L_{i'}(x, \vec x')$,  is crucial to our
work.  Therefore we obtain this structure carefully starting from
the bulk propagator $G_{\m \n'}(x,x')$ which we  derive. We begin
with a brief discussion of scalar fields.

\subsection{Scalar propagators}

For a scalar field in Euclidean $AdS_{d+1}$, its bulk-to-boundary propagator $K_\D(x,\vec x')$ can be obtained from the bulk propagator $G_\D(x,x')$ in the following limit:
\be
K_\D(x,\vec x') = \lim_{x_0'\to0} (2\D-d) x_0'^{-\D} G_\D(x,x') \,.
\ee
Explicitly, the scalar bulk-to-boundary propagator is
\be\label{btbdy}
K_\D(x,\vec x') = C_\D \[\fr{x_0}{x_0^2 +(\vec x -\vec x')^2}\]^\D \,, \qquad\quad C_\D=\frac{\G(\D)}{\pi^{d/2}\G(\D-d/2)} \,.
\ee
It satisfies the equation of motion $(\Box - m^2) K_\D(x,\vec x') =0$, and the  boundary condition:
\be\label{bdylim}
\lim_{x_0\to0} x_0^{\D-d} K_\D(x,\vec x') = \d^{(d)}(\vec x -\vec x') \,.
\ee
The two-point function of the dual CFT operator is \cite{Freedman:1998tz}:
\be
\<\co_\D(\vec x)\co_\D(\vec x')\> = \lim_{x_0\to0} (2\D-d) x_0^{-\D} K_\D(x,\vec x') = \fr{(2\D-d)C_\D}{(\vec x -\vec x')^{2\D}} \,.
\ee

\sse{Bulk propagator for the Chern-Simons gauge field}

The bulk propagator $G_{\m\n'}(x,x')$ of an abelian Chern-Simons  gauge field
must produce solutions of the inhomogeneous equation
\be \label{cswsource}
\e^{\r\m\n} \pa_\r A_\m = -\sqrt{g} J^\n
\ee
with a conserved source current $J^\n(x)$ in the bulk.  The solution
\be \label{intsol}
A_\m = \int d^3x' \sqrt{g(x')} G_{\m\n'} J^{\n'}(x')
\ee
suggests the naive equation of motion 
\be
\e^{\r\m\n} \pa_\r G_{\m\n'}(x,x') = -\d^\n_{\n'} \d(x,x') \,,
\ee
in which $\e^{\r\m\n}$ and $\d(x,x')$ transform as tensor densities and $G_{\m\n'}$ is a bi-vector.  However, this equation is inconsistent because the gauge invariant differential operator is not invertible.
Therefore we follow \cite{D'Hoker:1999jc} and use the modified equation
\be\label{geq}
\e^{\r\m\n} \pa_\r G_{\m\n'}(x,x') = -\d^\n_{\n'} \d(x,x') +\sqrt{g} \pa_{\n'} \Omega^\n(x,x') \,.
\ee
The solution (\ref{intsol}) remains valid since the current is conserved.

The most general $SO(3,1)$ invariant ansatz for $G_{\m\n'}$ is
\be
G_{\m\n'} = -(\pa_\m \pa_{\n'} u) F(u) + \pa_\m \pa_{\n'} S(u) +
\sqrt{g}\e_{\m\r\s} (\pa^\r \pa_{\n'} u) (\pa^\s u) T(u) \,,
\ee
in which $u$ is the bi-invariant variable
\be\la{udef}
u \eq \fr{(x-x')^2}{2x_0 x_0'} \,.
\ee
Similarly the most general ansatz for $\Omega^\n$ is
\be
\Omega^\n = (\pa^\n u) \Omega(u) \,.
\ee
We substitute this
ansatz in (\ref{geq})  and  use (2.9)-(2.15) of \cite{D'Hoker:1999jc}. For
$x \ne x'$,  the coefficients of the independent bi-vectors
$D^\m\pa_{\n'}u$ and $D^\m u\pa_{\n'} u$ give the differential
equations
\ba
F' &= 0 \,,\\
u(u+2) T' +2(u+1) T &= \Omega \,,\\
-(u+1) T' -2 T &= \Omega' \,.
\ea
Therefore $F$ is a constant and can be absorbed into $S$.  The last
two equations give
\be
u(u+2) T'' +5(u+1) T' +4T =0 \,,
\ee
from which we find
\be
T(u) = \fr{u+1}{\[u(u+2)\]^{3/2}}
\ee
up to an overall constant.  Setting $S=0$, we find the bulk propagator
\be\label{bulkprop}
G_{\m\n'} = \sqrt{g(x)}\e_{\m\r\s} (\pa^\r \pa_{\n'} u) (\pa^\s u) T(u) \,.
\ee

This propagator satisfies the Lorentz  gauge condition in both variables, i.e. $D^\m G_{\m\n'}=0$ and $D^{\n'} G_{\m\n'}=0$.  As written above the propagator is not normalized.  It can be shown that  $G_{\m\n'}/(4\pi)$ satisfies \er{geq} with the correct coefficient of the $\d$-function.  The normalized form is not needed in this paper.

\sse{Bulk-to-boundary propagator}

We define the bulk-to-boundary propagator (up to an overall constant) as
\be
K_{\m i'}(x,\vec x') = \lim_{x_0'\to0} G_{\m i'}(x,x') \,,
\ee
from which we find
\be
K_{0 i'} = - \fr{4x_0 \e_{0i'j} (x-x')^j}{\[x_0^2 +(\vec x -\vec x')^2\]^2} = \pa_0 \[\fr{2\e_{0i'j} (x-x')^j}{x_0^2 +(\vec x -\vec x')^2}\] \,,
\ee
and
\be  \label{kiip}
K_{i i'} = \fr{4}{\[x_0^2 +(\vec x -\vec x')^2\]^2} \[\fr{\e_{0 i' i}}{2} \[x_0^2 -(\vec x -\vec x')^2\] -\e_{0 i j} (x-x')_{i'} (x-x')^j\] \,.
\ee
Using the identity
\be
\e_{0ij} V_{i'} V^j = \e_{0i'j} V_i V^j - \e_{0i'i} \vec V^2
\ee
which may be verified explicitly for an arbitrary vector $\vec V$, we can rewrite $K_{ii'}$ as
\ba
K_{i i'} &= \fr{4}{\[x_0^2 +(\vec x -\vec x')^2\]^2} \[\fr{\e_{0 i' i}}{2} \[x_0^2 +(\vec x -\vec x')^2\] -\e_{0 i' j} (x-x')_i (x-x')^j\] \\
&= \pa_i \[\fr{2\e_{0i'j} (x-x')^j}{x_0^2 +(\vec x -\vec x')^2}\] \,.
\ea
Thus $K_{\m i'}$ is a ``pure gauge,'' specifically
\be\label{puregauge}
K_{\m i'}(x,\vec x') = \frac{\pa}{\pa x^\m} \L_{i'} \,, \qquad \qquad
\L_{i'} = \frac{2\e_{0i'j} (x-x')^j}{x_0^2 +(\vec x -\vec x')^2} \,.
\ee
This is not surprising since $F_{\m\n} =0$ when there is no bulk current source.

It is curious to observe, from (48) of \cite{Freedman:1998tz}, that the (normalized) bulk-to-boundary propagator for a Maxwell gauge field in AdS$_3$ is also a pure gauge, namely
\be
K_{\m i'}^{\rm Maxwell}(x,\vec x')= \frac{1}{2\pi}\pa_\m \[\fr{(x-x')_{i'}}{x_0^2 +(\vec x -\vec x')^2}\]\,.
\ee

\subsection{Bulk-to-boundary propagator in holomorphic components}

We briefly state conventions for holomorphic components in the 2-plane initially described by Cartesian coordinates $z_1,~z_2$ with metric $\d_{ij}.$
\bea \label{holocoords}
z &=& z_1+i\,z_2 \,,\qquad \quad \bar z = z_1-i\,z_2 \,,\\
g_{z\bar z} &=& g_{\bar z z } = 1/2 \,,\qquad \quad  g_{zz} = g_{\bar z\bar z} = 0 \,,\\
g^{z\bar z} &=& g^{\bar z z} = 2 \,,\qquad \quad  g^{zz} = g^{\bar z\bar z} = 0 \,,\\
\e_{z\bar z} &=& -\e_{\bar z z} =i/2 \,,\qquad \quad  \e^{z\bar z} = -\e^{\bar z z} = -2i \,.
\eea
Note that the alternating symbol is defined by $\e_{12}=\e^{12}=1$ in Cartesian coordinates and transformed as a tensor to holomorphic coordinates.
Note also that $\sum_i z_iz_i = z\bar z= |z|^2$.  The holomorphic components of 1-forms are given by $A_z = (A_1-i A_2)/2$, $A_{\bar z} = (A_1+iA_2)/2$.
Similarly $\pa_z = (\pa_1 -i \pa_2)/2$, $\pa_{\bar z} =(\pa_1 + i \pa_2)/2$.
\vspace{3mm}

%
In many calculations it is convenient to use holomorphic and anti-holomorphic components of the bulk-to-boundary propagator.
We use $x^0,~x,~\bar x$ for the bulk point and $z,~\bar z$ for the boundary point. In the conventions of (\ref{holocoords}),  we have
\be\la{holol}
\L_z = i\frac{\bar x - \bar z}{x_0^2 + |x-z|^2} \,,\qqu\qqu
\L_\zb = -i\frac{x - z}{x_0^2 + |x-z|^2} \,,
\ee
and the  holomorphic and anti-holomorphic components of (\ref{puregauge}) are
\bea\label{holoprops}
K_{x z} &=& i \pa_x \bigg[\frac{\bar x - \bar z}{x_0^2 + |x-z|^2}\bigg] \,,\qquad\qquad K_{\bar x \bar z} =-i \pa_{\bar x} \bigg[\frac{x - z}{x_0^2 + |x-z|^2 }\bigg] \,,\\
K_{\bar x z} &=& i\pa_{\bar x} \bigg[\frac{\bar x - \bar z}{x_0^2 + |x-z|^2}\bigg] \,, \qquad\qquad K_{ x \bar z} =-i \pa_{x} \bigg[\frac{x -  z}{x_0^2 + |x-z|^2 }\bigg] \,,\\
K_{0 z} &=& i\pa_0 \bigg[\frac{\bar x - \bar z}{x_0^2 +
|x-z|^2}\bigg] \,,\qquad\qquad K_{0 \bar z} =-i \pa_{0} \bigg[\frac{x -
z}{x_0^2 + |x-z|^2 }\bigg]\,. \eea

We write the formal limit $x_0\to 0$ of these propagators as
\bg\label{lim}
K_{xz} \to -i \frac{1}{(x-z)^2} \,,\qquad\quad
K_{\bar x z} \to i \pi \d^{(2)}(\vec x -\vec z) \,,\\
K_{\bar x \bar z} \to i \frac{1}{(\bar x-\bar z)^2} \,,\qquad\quad
K_{x \bar z} \to -i \pi \d^{(2)}(\vec x -\vec z) \,.
\eg
Note that these are indeed the desired boundary conditions for the bulk-to-boundary propagators of Chern-Simons gauge fields $A$ and $\td A$.  In order to see this, we recall that the Chern-Simons gauge fields satisfy first-order equations of motion in the bulk.  Therefore, a consistent boundary condition can only be imposed on half of the two boundary components of $A$ or $\td A$.  Since $A$ is dual to a holomorphic current $J_z$ in the CFT, we should impose a Dirichlet boundary condition on $A^z$, i.e. we specify the value $A^z \to A^z_\partial$ on the boundary.   The correct normalization is given by the Euclidean AdS/CFT dictionary
\be
Z_{bulk}[A_\pa^z] = Z_{CFT}[A_\pa^z] \eq \< e^{2\pi i\int A_\pa^z J_z} \>_{CFT} \,,
\ee
where $A_\pa^z$ is the boundary value of $A^z$, and the prefactor of $2\pi i$ is consistent with the conventional normalization for a holomorphic current in a two-dimensional CFT.  The bulk-to-boundary propagator $\T Kxz = 2K_{\bar x z}$ is responsible for constructing a bulk solution $A^\m$ from the source of the boundary current $J_z$ (which is $2\pi i A_\pa^z$), and the normalization in (\re{lim}) is precisely what we need.  A similar argument holds for $\td A$.

\section{Appendix: Holomorphic $\<JJ\>$ and $\<JJJ\>$ in the undeformed CFT}

The purpose of this appendix is to describe the use of the bulk-to-boundary
propagators to calculate the correlation functions $\<J(y)J(z)\>$
and $\<J(y)J(z)J(w)\>$ in which the holomorphic components of
conserved currents appear.   For non-abelian
currents,  the bulk calculations can be compared with the result of  OPE methods  in the dual CFT.
This provides a
test of the normalization of the bulk-to-boundary propagator.

\subsection{SU(2) Chern-Simons action}

The normalized Euclidean Chern-Simons action for the group SU(2) and level $k$ is\footnote{The bulk term agrees with \cite{Witten:1988hf} and the boundary term is taken from \cite{Elitzur:1989nr, Aharony:2001dp}.}
\bea
S &=& \frac{k}{8\pi} \int_{\text{bulk}} \(A^a\wedge dA^a +\fr{1}{3} \e^{abc} A^a \wg A^b \wg A^c\)\,-\, \frac{ik}{16\pi}\int_{\text{bdy}}A^a\wedge *A^a\\
&=&
\frac{k}{8\pi} \int_{\text{bulk}} d^3x\, \e^{\m\r\n}A_\m^a \( \pa_\r A_\n^a +\frac13 \e^{abc}A_\r^bA_\n^c\) - \frac{ik}{16\pi}\int_{\text{bdy}} d^2w \[(A_1^a)^2 + (A_2^a)^2\]\,.
\eea
The purpose of the boundary term is to enforce the condition that $A_{\bar w}$ vanishes on the boundary.  We will achieve this by the dropping the propagators $K_{\mu \bar w}$ with anti-holomorphic boundary index. Note that all three  components  of $K_{\mu \bar w} $ are non-vanishing in the bulk.  The generator of AdS/CFT correlators is actually $e^{iS}$ (instead of the usual $e^{-S}$ in the Euclidean signature) because the Chern-Simons action does not change by the factor $i$ under Wick rotation.  We will therefore insert an extra factor of $i$ in the results for the two-point and three-point functions below.

\subsection{$\<J(y)J(z)\>$}
To calculate $\<J^a(y)J^b(z)\>$ we reexpress the boundary action as
$S_{\text{bdy}} = -\frac{ik}{4\pi} \int_{\text{bdy}} A_w^a A_{\bar w}^a\,.$  We regulate the resulting Witten diagram  by assuming initially that
the boundary integral is evaluated at the small radial coordinate value $w_0$ with subsequent limit $w_0\to 0$.
The diagram contains two ``Wick contractions'' and thus produces
\be \label{jj}
\<J^a(y) J^b(z)\> = -i\d^{ab} \fr{ik}{4\pi}\int d^2w\bigg[K_{wy}(w_0, \vec{w}-\vec{y})  K_{\bar w z}(w_0, \vec{w}-\vec{z})+ (y \leftrightarrow z)\bigg] \,,
\ee
where we have inserted an extra factor of $i$ as mentioned above.
Using \er{lim} we see that the  formal limit $w_0\to 0$ of the integral gives the holomorphic correlator\footnote{The limit $w_0\to 0$ of an  analytic evaluation of the regulated integral using Feynman parameters gives the same result. }
\be \label{2ptj}
\<J^a(y) J^b(z)\> = \frac{k}{2}\,\frac{\d^{ab}}{(y-z)^2}\,.
\ee
This agrees with the  result for the ${\cal N} =4$ CFT containing $2k$ complex scalars and $2k$ Dirac spinors.

\subsection{$\<J^a(y)J^b(z)J^c(w)\>$}


The 3-point function is given by the Witten diagram with the cubic vertex from \er{csaction} with  three  bulk-to-boundary propagators.
Counting 6 Wick contractions, the 3-point function is given by the
integral: \be \<J^a(y)J^b(z) J^c(w)\> = \frac{ik}{4\pi}
\e^{abc} \int d^3x \e^{\m\n\r} \bigg[\pa_\m
\L_y(x,y) \pa_\n\L_z(x,z)
\pa_\r\L_w(x,w)\bigg] \,, \ee
where we have inserted an extra factor of $i$, and $y,z,w$ denote holomorphic  components of $\L_i$.  We integrate $\pa_\m$ by parts.  It is immediately obvious that the resulting bulk integral vanishes, but it leaves the  boundary integral
\be \<J^a(y)J^b(z) J^c(w)\> = \frac{ik}{4\pi} \e^{abc} \lim_{x_0\to 0} \int
d^2x \e^{0\n\r}
\L_y(x,y) \pa_\n\L_z(x,z)
\pa_\r\L_w(x,w)
\ee
We now substitute the appropriate $\L$ factors from \er{holol}, and use
$\e^{0 w \bar w}=-2i $ to write
 \bea
 \<J^a(y)J^b(z)J^c(w)\> &=&
 -\frac{ik}{2\pi} \e^{abc}\lim_{x_0\to 0} \int d^2x\frac{\bar x - \bar y}{x_0^2 + |x-y|^2}\bigg( \pa_x\bigg[\frac{\bar x - \bar z}{x_0^2 + |x-z|^2}\bigg] \nonumber\\
&&\ \ \ \ \ \ \ \ \ \ \ \times\pa_{\bar x} \bigg[ \frac{\bar x -
\bar w}{x_0^2 + |x-w|^2}\bigg] - z \leftrightarrow w\bigg) \eea
The formal limit of this expression, obtained from \er{lim}, is
\be\label{3pt}
 \<J^a(y)J^b(z)J^c(w)\> = \frac{ik}{2} \frac{\e^{abc}}{(w-z)^2}\bigg( \frac{1}{w-y} - \frac{1}{z-y}\bigg) =- \fr{ik}{2} \frac{\e^{abc}}{(y-z)(z-w)(w-y)}\,.
\ee
This is the  correct form of the 3-point function.   The result should be multiplied by $i$ as discussed above.

\sse{Compatibility with the OPE}

In the free CFT with $k$ complex scalars and $k$ Dirac fermions transforming in the fundamental representation of SU(2), the
SU(2) R-current is $J^a = \sum_{i=1}^k (\bar\psi_i \tau^a \psi_i)/2$ where
$\tau^a$ are the three Pauli matrices.  The basic OPEs are \be
\label{opes} \bar\psi_i(z) \psi_j(0) \sim \frac{\d_{ij}}{z} \,,\qquad\qquad J^a(z)
J^b(0) \sim \frac{k}{2z^2} \d^{ab} +\fr{i}{z} \e^{abc} J^c(0)\,. \ee
From this one can quickly write the 2-point correlator as  $\<J^a(y)
J^b(z)\> = \frac{k}{2}\frac{\d^{ab}}{(y-z)^2}$  which agrees with \er{2ptj}. To check
\er{3pt}  we take the limit $y \to z$.  In this limit \er{opes} requires
\be
\<J^a(y)J^b(z)J^c(w)\> \to \frac{i \e^{abd}}{y-z} \<J^d(z)
J^c(w)\> = \fr{i k}{2} \fr{\e^{abc}}{(y-z)(z-w)^2}\,,
\ee
which is indeed satisfied by \er{3pt}.

\se{Existence of cubic coupling}\label{sec:Cubic}

In this paper, several calculations are based on the existence of
cubic coupling. Although our results can be generalized to higher
derivative vertices, it would be nice if there is a concrete example
to show that the existence of cubic coupling is consistent with
supergravity in $AdS_3$. In the following, we are going to show that
the cubic coupling can be very naturally generated from Kahler
potential.

Let us start from the Kahler potential,
\begin{eqnarray}
  \label{eq:Kahler}
K&=& \Phi^\dag\Phi+Z^\dag Z\(1+\frac{\lambda}{M} (\Phi+\Phi^\dag)\) \,.
\end{eqnarray}
Here $Z$ is the chiral supermultiplet with non-zero $R$-charge, and
$\Phi$ is taken to be neutral under $R$-symmetry, and its scalar
component is the moduli field. $M$ is the suppression scale of the
irrelevant operator. One can write the scalar part of the Lagrangian
induced by the Kahler potential as
\begin{eqnarray}
  \label{eq:CubLag}
\mathscr{L}&\supset& \partial_\mu \phi^\dag \partial^\mu \phi+D_\mu
z^\dag D^\mu z + \lambda(\phi D_\mu z^\dag D^\mu z+\phi^\dag D_\mu
z^\dag D^\mu
z)\nonumber\\
&&\qqu+\lambda(\partial_\mu\phi D^\mu z^\dag\  z+\partial_\mu\phi^\dag
 z^\dag  D^\mu z)+...
\end{eqnarray}
where $(...)$ denotes the rest of the Lagrangian. Integrating by
parts on the last two terms gives
\begin{eqnarray}
  \label{eq:CubLag2}
\mathscr{L}&\supset& \partial_\mu \phi^\dag \partial^\mu
\phi+\partial_\mu z^\dag \partial^\mu z -\lambda \phi  \Box z^\dag
z-\lambda \phi^\dag z^\dag \Box z+...
\end{eqnarray}
Now let us prove $-\lambda \phi^\dag z^\dag \Box z$ can be replaced
by a simple cubic term using the equation of motion of $z$.
According to \cite{Weinberg:1995mt}, a coupling constant is
redundant if the variation of such coupling constant vanishes when
we use field equation of motion. The field equation for $z$ can be
written as
\begin{eqnarray}
  \label{eq:EOMZ}
\Box z-m^2 z + f(z,\phi)=0
\end{eqnarray}
where $m$ is the mass of $z$. $f(z,\phi)$ is the nonlinear terms
from the interactions of the Lagrangian. Since we only focus on
cubic vertices, the explicit forms of those terms are not important.
Then we can add an additional term to Lagrangian with an arbitrary
coupling $\lambda'$ to Eq.~(\ref{eq:CubLag2}), and get
\begin{eqnarray}
  \label{eq:CubLag3}
\mathscr{L}&\supset& \partial_\mu \phi^\dag \partial^\mu \phi+D_\mu
z^\dag D^\mu z -\lambda \phi  \Box z^\dag z-\lambda \phi^\dag z^\dag
\Box z +\lambda'\phi z^\dag(\Box z-m^2 z
+ f(z,\phi))+...\nonumber\\
\end{eqnarray}
Taking $\lambda'=\lambda$, we see $(-\lambda \phi^\dag z^\dag \Box
z)$ is replaced by mass term of $z$ plus vertices with higher order
of fields, i.e.
\begin{eqnarray}
  \label{eq:CubLag4}
\mathscr{L}&\supset& \partial_\mu \phi^\dag \partial^\mu \phi+D_\mu
z^\dag D^\mu z -\lambda \phi  \Box z^\dag\ z +\lambda\phi
z^\dag(-m^2 z + f(z,\phi))+...
\end{eqnarray}
Similarly, one can apply the equation of motion for $z^\dag$ to replace
$(-\lambda \phi  \Box z^\dag\ z)$ by $(\lambda m^2 \phi z^\dag z)$
plus vertices with higher order field dependence. Now we see the
existence of cubic couplings is quite generic.

\end{document}